\documentclass{aa}
\usepackage{graphicx}
\usepackage{txfonts}
\usepackage{hyperref}
\hypersetup{
    colorlinks=true,
    linkcolor=blue,
    filecolor=magenta,      
    urlcolor=blue,
    citecolor=blue,
    pdfpagemode=FullScreen,
    }
\usepackage{graphicx}	% Including figure files
\usepackage{amsmath}	% Advanced maths commands
\usepackage{caption}
\usepackage{subcaption}
\usepackage{multirow}
\usepackage{pifont}% http://ctan.org/pkg/pifont
\usepackage[normalem]{ulem}
\usepackage{booktabs}
\newcommand{\mass}{log(M_{\star}/M_{\odot})}

\newcommand{\kmsmpc}{km\:s^{-1}\:Mpc^{-1}}

\defcitealias{Bruzual2003}{BC16}
\defcitealias{Goncalves2020}{G20}
\defcitealias{CabreraZiri2022}{CC22}

\begin{document}

   \title{Globular clusters as cosmic clocks: new cosmological hints from their integrated light}

   \author{Elena Tomasetti\inst{1,2} \fnmsep\thanks{\email{elena.tomasetti2@unibo.it}}
          \and
           Michele Moresco\inst{1,2}
           \and
           Carmela Lardo\inst{1,2}
           \and 
           Andrea Cimatti\inst{1,3}
           \and
           Raul Jimenez\inst{4,5}
           }

   \institute{
            Dipartimento di Fisica e Astronomia “Augusto Righi”–Università di Bologna, via Piero Gobetti 93/2, I-40129 Bologna, Italy
            \and
            INAF - Osservatorio di Astrofisica e Scienza dello Spazio di Bologna, via Piero Gobetti 93/3, I-40129 Bologna, Italy
            \and
            INAF - Osservatorio Astrofisico di Arcetri, Largo E. Fermi 5, I-50125, Firenze, Italy
            \and
            ICC, University of Barcelona, Mart\' i i Franqu\` es, 1, E08028, Barcelona, Spain
            \and
            ICREA, Pg. Lluis Companys 23, Barcelona, 08010, Spain
             }

   \date{October 29, 2024}

    \titlerunning{Globular clusters as cosmic clocks}
    \authorrunning{E. Tomasetti et al.}
% \abstract{}{}{}{}{} 
% 5 {} token are mandatory
 
  \abstract
   {}
   {In this work we explore the reliability and robustness in measuring the ages and main physical properties of a sample of old Milky Way globular clusters (GCs) from their integrated light. This approach sets the stage for using GCs as cosmic clocks at high redshift. Additionally, it enables us to establish an independent lower limit to the age of the Universe, and an upper limit to $H_0$.}
   {We analyse a sample of 77 GCs from the WAGGS project, by first measuring their spectral features (Lick indices and spectroscopic breaks) with \texttt{PyLick} and then performing full-spectral-fitting with \texttt{BAGPIPES}. The analysis of Lick indices offers an initial estimate of the population's age and metallicity, generally aligning well with values reported in the literature. However, it also highlights a subset of old clusters for which we estimate younger ages. This discrepancy is primarily attributed to the presence of horizontal branches with complex morphologies, which are not accounted for in the stellar population models. With full-spectral-fitting we measure the GCs' ages, metallicities, and masses, testing how removing the cosmological prior on the ages affects the final results.}
   {Compared to isochrone fitting estimates, ages are best recovered when the cosmological prior is removed, with a 20\% increase in the number of GCs showing ages compatible with literature values within $\pm$1.5 Gyr. The derived metallicity and mass are consistently in good agreement with the reference values, regardless of HB morphology, [Z/H], or the fit settings. The average discrepancies across the entire sample are $\mathrm{\langle \Delta [Z/H] \rangle = -0.02 \pm 0.24}$ dex for metallicity and $\mathrm{\langle \Delta \mass \rangle = 0.04 \pm 0.28}$ dex for mass. Metal-rich GCs ([Z/H]$\geq$-0.4) showing a red horizontal branch (HBR>0) are the sub-group in which ages are best recovered. In this group, 70\% of the results align with literature values within $\pm$1.5 Gyr. Identifying the tail of the oldest cosmology-independent ages with a Gaussian Mixture Model, we obtain a sample of 24 objects with $\mathrm{\langle age \rangle = 13.4 \pm 1.1 \:Gyr}$.}
   {Being a natural lower limit to the age of the Universe, we use the age of the oldest GCs to constrain the Hubble constant, obtaining $\mathrm{H_0 = 70.5^{+7.7}_{-6.3}\: \kmsmpc}$ (stat+syst) when a flat $\Lambda$CDM with $\mathrm{\Omega_m = 0.30 \pm 0.02}$ (based on low-z measurements) is assumed. Validating the analysis of GCs based on their integrated light lays the foundation to extend this type of study to high redshift, where GCs have begun to appear in lensed fields, thanks to JWST.}


   \maketitle
%
%-------------------------------------------------------------------

\section{Introduction}
\label{sec:1Intro}
During the last decades, the flat $\Lambda$CDM model has been a fundamental pillar in cosmology thanks to the support of independent cosmological probes, including the cosmic microwave background \citep[CMB; e.g. ][]{Bennett2003,PlanckCollaboration2020}, type Ia supernovae \citep[e.g. ][]{Riess1998,Perlmutter1999}, and baryon acoustic oscillations \citep[e.g.][]{Percival2001,Eisenstein2005}. Nonetheless, further investigation is necessary to unveil the nature of dark matter and dark energy and assess the precise values of cosmological parameters. In fact, due to the higher precision achieved in late- and early-Universe probes, some inconsistencies have emerged regarding the value of the Hubble constant ($H_0$), where a tension of 4-5$\sigma$ has now been observed \citep{Abdalla2022}.   

In this context, the age of the Universe ($t_U$) can play a crucial role, given its sensitivity to $H_0$. Indeed, in a flat $\Lambda$CDM cosmology with $\Omega_M$=0.3 and $\Omega_\Lambda$=0.7, $t_U$ can span a range from $\sim$14.1 Gyr if $H_0$=67 $\mathrm{\kmsmpc}$ to $\sim$12.9 Gyr if $H_0$=73 $\mathrm{\kmsmpc}$.
Thus, measuring the absolute ages of the most long-lived objects at $z=0$ can be critical since they naturally place a lower limit on the current age of the Universe ($t_U$) and, in turn, an upper limit on $H_0$. This provides independent constraints on the Hubble constant and valuable information for investigating the origin of the observed tension. 

Globular clusters (GC) are among the oldest objects in the Universe for which we can accurately determine the age \citep{VandenBerg1996,Soderblom2010,Brown2018,Oliveira2020}. Composed of roughly one million stars that formed simultaneously with similar composition \citep[though see reviews by][for discussions on the multiple population phenomenon in massive clusters]{Bastian2018,Gratton2019,Milone2022}, these clusters have remained gravitationally bound for up to a Hubble time. Each GC thus serves as an observable record of the age, metallicity, and kinematics from the time of its formation. Therefore, by measuring their ages, we can use them as "clocks" that began ticking in the early stages of the Universe's evolution \citep{OMalley2017, Jimenez2019, Valcin2020, Valcin2021,Cimatti2023}.

The most straightforward method to determine the age of a GC is by exploiting the fact that the position of the main-sequence turn-off (MSTO) in the plane of effective temperature (T$_{\rm eff}$) versus luminosity (L) changes with age (or mass). Isochrones, or theoretical tracks of stars with the same chemical composition, are fitted to the MSTO region of colour-magnitude diagrams (CMD) to estimate the age. Even if this is a well-established and robust method, it is important to explore new and complementary approaches that can address the case when the CMD is not available. In need of a spatially resolved stellar population, indeed, isochrone fitting can be applied only to nearby systems, while moving further than the Magellanic Clouds becomes either very expensive in terms of exposure time or even impossible. Moreover, recent JWST observations of lensed fields highlighted the presence of GC candidates around lensed galaxies, like the Sparkler \citep{Mowla2022} at z=1.38 which, if confirmed with spectroscopy, would extend the study of GCs at high redshift. To do so, we need to explore methods relying on GCs' integrated light and validate them against the traditional methods. In this scenario, one of the best ways to leverage all the integrated light information is to perform full-spectral-fitting (FSF), a technique that enables measuring, alongside the age, all the physical properties of the GC such as metallicity, mass, and dust reddening.

Previous works have derived physical parameters of GCs like age and metallicity using the integrated light provided by \citet{Schiavon2005} for 41 MW GCs \citep[e.g.,][]{Koleva2008,Cezario2013,CabreraZiri2022}, testing different algorithms, like \texttt{STECKMAP} \citep{Ocvirk2006}, \texttt{NBURSTS} \citep{Chilingarian2007}, \texttt{ULySS} \citep{Koleva2009} or \texttt{ALF} \citep{Conroy2012}, and different simple stellar population (SSP) models \citep{Bruzual2003,Prugniel2004,Vazdekis2010,Vazdekis2015}.  Others have benefited from the larger spectral coverage and higher resolution of the WiFeS Atlas of Galactic Globular cluster Spectra project \citep[WAGGS,][]{Usher2017,Usher2019a}, providing integrated spectra for 113 GCs in the MW and its satellite galaxies \citep{Usher2019,Goncalves2020,CabreraZiri2022}. In \citet{Goncalves2020}, for instance, the authors adopted the non-parametric FSF code \texttt{STARLIGHT} \citep{CidFernandes2005}, relying on MILES SSP models \citep{Vazdekis2015}, with a focus on how the wavelength range influences the recovery of the stellar parameters compared to the CMD fitting. \citet{CabreraZiri2022} extended the Milky Way GCs sample from \citet{Schiavon2005} by including younger objects from the Large and Small Magellanic Clouds (LMC and SMC) with WAGGS spectra, adopting the non-parametric FSF code \texttt{ALF}. While there is broad agreement that the ages of younger GCs can be reliably determined through FSF, these studies highlighted the challenges in dating the oldest GCs from their integrated spectra, often yielding results significantly younger than those from isochrone-fitting methods.

In this paper we focus on the oldest tail of the WAGGS GCs, analysing 82 GCs in the Milky Way. We take advantage of the high-quality integrated spectra provided by WAGGS, along with the wealth of data available for these objects, the independent age estimates derived with different techniques, and the fact that GCs are among the simplest stellar systems in the Universe, the closest templates to an SSP we have. In this study, we adopt a parametric FSF method, enabling the reconstruction of GCs' integrated emission within a high-dimensional parameter space. For this purpose, we use the FSF code \texttt{BAGPIPES} \citep{Carnall2018} utilizing the 2016 version of the \citet{Bruzual2003} SSP models. Previous studies in the literature have typically derived parameters such as age, metallicity, mass, and dust reddening while assuming a cosmological prior on age. In contrast, the novelty of this study lies in removing this prior to explore how the derived ages are affected, as done by \citet{Tomasetti2023} and \citet{Jiao2023}, in order to test the potential of the results in a cosmological framework. By testing this approach, we aim to assess its potential in a cosmological context. We also use these cosmology-independent results to place new constraints on $H_0$ setting the stage for future applications in studying the distant Universe.

This paper is organised as follows: in Sect. \ref{sec:2DATA} we describe the spectra we used, along with the adjustments needed and the ancillary data; in Sect. \ref{sec:3SPEC_ANALYSIS} the spectroscopic analysis of the sample is presented; in Sect. \ref{sec:4METHOD} the FSF method and its result are outlined; in Sect. \ref{sec:5COSMO} we report the final cosmological analysis; in Sect. \ref{sec:6CONCLUSIONS} we draw our conclusions.

%##############se#######  DATA #####################

\section{Data}
\label{sec:2DATA}

The WAGGS project \citep{Usher2017} is a library of integrated spectra of GCs in the Milky Way (MW) and the Local Group, obtained with the WiFeS integral field spectrograph on the Australian National University 2.3 m telescope. With 112 spectra of GCs in the Local Group, it is one of the largest GCs spectral libraries currently available, with a wide wavelength coverage (3270-9050 $\AA$) and high spectral resolution ($\mathrm{R \sim 6800}$). The spectra we work with are normalised and consist of four different gratings, each with its own sampling: 3270-4350 $\AA$ (0.27 $\AA$ per pixel), 4170-5540 $\AA$ (0.37 $\AA$ per pixel), 5280-7020 $\AA$ (0.44 $\AA$ per pixel) and 6800-9050 $\AA$ (0.57 $\AA$ per pixel). To perform FSF across the entire spectrum, certain adjustments were required.

First, we had to re-scale each spectrum to match its literature photometry, to retrieve the fluxes in physical units. We used the UBVRI integrated photometry from the 2010 edition of the Harris catalogue \citep{Harris1996,Harris2010}. The correction factor $C$, derived via $\chi^2$ minimisation, can be written as:
\begin{equation}
    \mathrm{C = \frac{\sum (p_J/e_J)^2}{\sum f_J p_J/e_J^2}},
\end{equation}
where $p$ is the photometry in the $J$-th filter, and $f$ and $e$ are the average flux and corresponding error estimated on the spectrum on a window of 10 $\AA$. We then multiplied the spectrum in each grating by the corresponding factor $C$.\\
Here, we must underline that UBVRI photometry is not available for all the objects in WAGGS, but only for the 82 GCs belonging to the MW. For the younger GCs in the LMC and SMC and in the Fornax dwarf spheroidal, only BV photometry is available, respectively from \citet{Bergh1981} and \citet{Bergh1969}. Anyway, in this work, we want to focus on the oldest tail of the local GCs, so we limit our sample to the MW GCs. Before proceeding with the analysis we performed a visual inspection of the spectra, removing five GCs showing either visibly corrupted or very noisy regions (S/N<10 in more than 40\% of the spectrum), namely NGC6144, NGC6401, NGC6517, NGC6712 and NGC7492. The sample we analyse here is then constituted of 77 GCs. 
 
To combine the four gratings into a single spectrum, we interpolated all of them onto a common wavelength grid, matching the largest spectral sampling value (0.57 $\AA\:$ per pixel). In the overlap regions, the flux and associated error were estimated by averaging the spectra from the consecutive gratings.

Throughout the paper, we compare our results to literature values of age, mass, and metallicity, and we also consider additional quantities to complement and expand our analysis, like dust reddening, radial velocities and distances of the GCs. We use as a reference the values listed in \citet{Usher2017} for ages and masses, in \citet{Harris2010} for metallicities ([Fe/H]) and dust reddening ($\mathrm{E_{B-V}}$) and in \citet{Baumgardt2023} for radial velocities, distance from the Sun and associated errors. 
\\As for the uncertainties on metallicities, we consider the errors found in other spectroscopic investigations based on integrated spectra of Galactic GCs, which are approximately $\pm$0.15 dex \citep[see][]{Roediger2014,Colucci2017}. 
\\On ages, the error budget based on MSTO fitting involves several key contributors. The most significant is distance uncertainty; an error of approximately 0.1-0.15 mag can result in about a 10\% uncertainty in age. The error in the initial helium content, known within $\sim$2\%, translates to about a 2\% uncertainty in age. An error in the global metallicity of $\sim$9-10\% and of $\sim$0.15 dex in iron content leads to approximately 4-5\% error in age. An uncertainty of $\sim$0.15-0.2 dex in alpha elements translates to about 4\% error in age \citep[see a discussion in][]{Cassisi2013}. Combining these factors, the overall uncertainty in age can be around 10-20\% \citep[e.g.][]{OMalley2017}. For the sake of comparison, we consider a fixed error of $\pm$1.5 Gyr. 
\\On mass, a typical uncertainty of $\sim$10\% is generally found on single measurements, corresponding to 0.05 dex in $\mathrm{\mass}$ \citep{HenaultBrunet2019}. Nevertheless, a difference of $\sim$0.2 dex on average can be observed among different catalogues (e.g., \citet{Usher2017} with respect to \citet{Baumgardt2023}). For this reason, we adopt a typical error on mass of 0.2 dex.

%%%%%%%%%%%%%%%  SPECTROSCOPIC ANALYSIS #########################
\section{Spectroscopic analysis}\label{sec:3SPEC_ANALYSIS}
Before estimating the physical parameters of our GC sample with FSF, we want to derive measurements of the spectroscopic features in our sample and use those to have a preliminary assessment of the age and metal content of our GCs. At high redshift, the study of Lick indices \citep{Burstein1984, Faber1985} or spectral breaks is often used to constrain stellar population properties. Here we want to see how GCs, the astrophysical objects that most resemble SSPs, fit inside this framework.

To do so, we first measured all the absorption features detectable in the spectra using the public code \texttt{PyLick} \citep{Borghi2022a}. We want to compare these data with theoretical models estimated by \cite{Thomas2011} at different ages, metallicities and alpha-enhancements ([$\alpha$/Fe]). Since these models are built with MILES resolution ($\sim$2.7 {\AA} FWHM), we downgraded the spectral resolution of our sample to match the models. Moreover, if an index lay in the overlap region of two spectral gratings, we ran \texttt{PyLick} on both spectra and then averaged the two values weighting them with their associated errors.

In particular, we measured indices of the Balmer series, like H$\beta$ and H$\gamma$, indices of the iron group like Fe5270 and Fe5335 or Mgb and broader features like the D4000, a spectral break at 4000 {\AA}. These features are particularly relevant in the study of stellar populations in that they are known to independently correlate with age (like D4000 and H$\beta$) and metallicity (like the iron group), so their analysis can give important insights into the physical properties of the GCs, as we discuss in the following sections.

\subsection{Index-age analysis}
After measuring the spectral features, we can compare their trend in age with the theoretical ones. A variety of stellar libraries are available in the literature, like \citet{Bruzual2003}, \citet{Maraston2011} and \citet{Conroy2012}, adopting different stellar evolutionary models, libraries of stellar spectra and procedures to compute the integrated spectra. In \citet{Moresco2012}, however, it is shown how the assumption of different SPS models has a negligible impact on the slope of the index-age trend, in particular in the case of the D4000. To compare our measurements with theoretical trends, here we are considering the 2016 version of \citet{Bruzual2003} models (hereafter BC16), since these are the same SPS models implemented in \texttt{BAGPIPES}. 

\begin{figure*}
    \centering
        \begin{subfigure}[b]{\textwidth}
         \centering
         \includegraphics[width=0.45\textwidth]{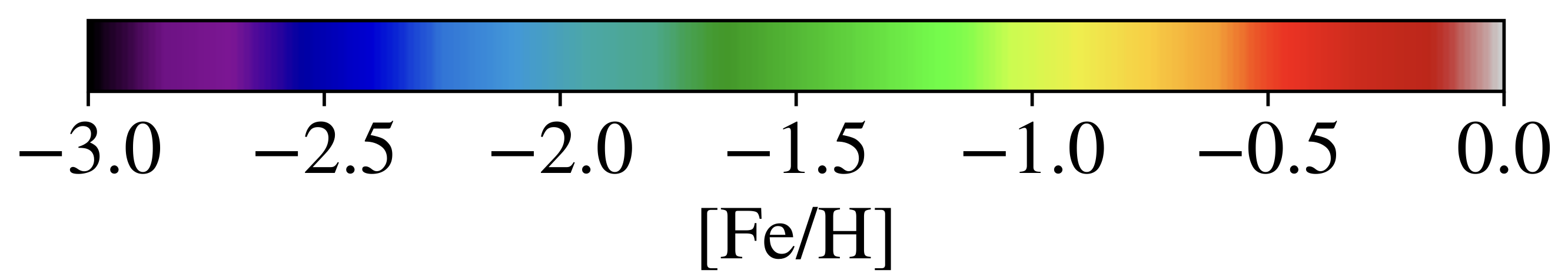} 
         \vspace{-0.1cm}
    \end{subfigure}
    \centering
        \begin{subfigure}[b]{0.98\textwidth}
         \centering
         \includegraphics[width=\textwidth]{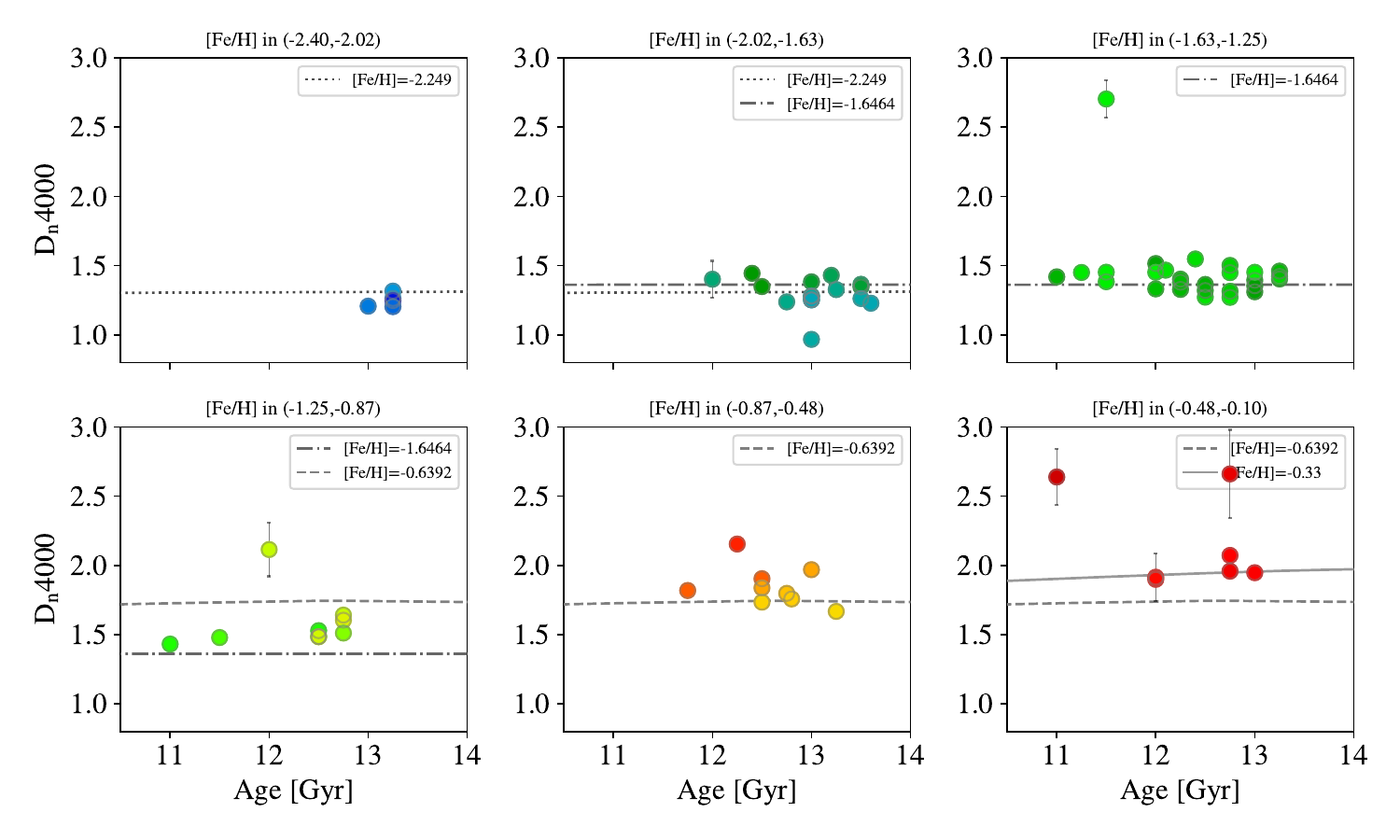}
         \vspace{-0.7cm}
         \caption{}
         \label{fig:D4000_age}
    \end{subfigure}
    \centering
        \begin{subfigure}[b]{0.98\textwidth}
         \centering
         \includegraphics[width=\textwidth]{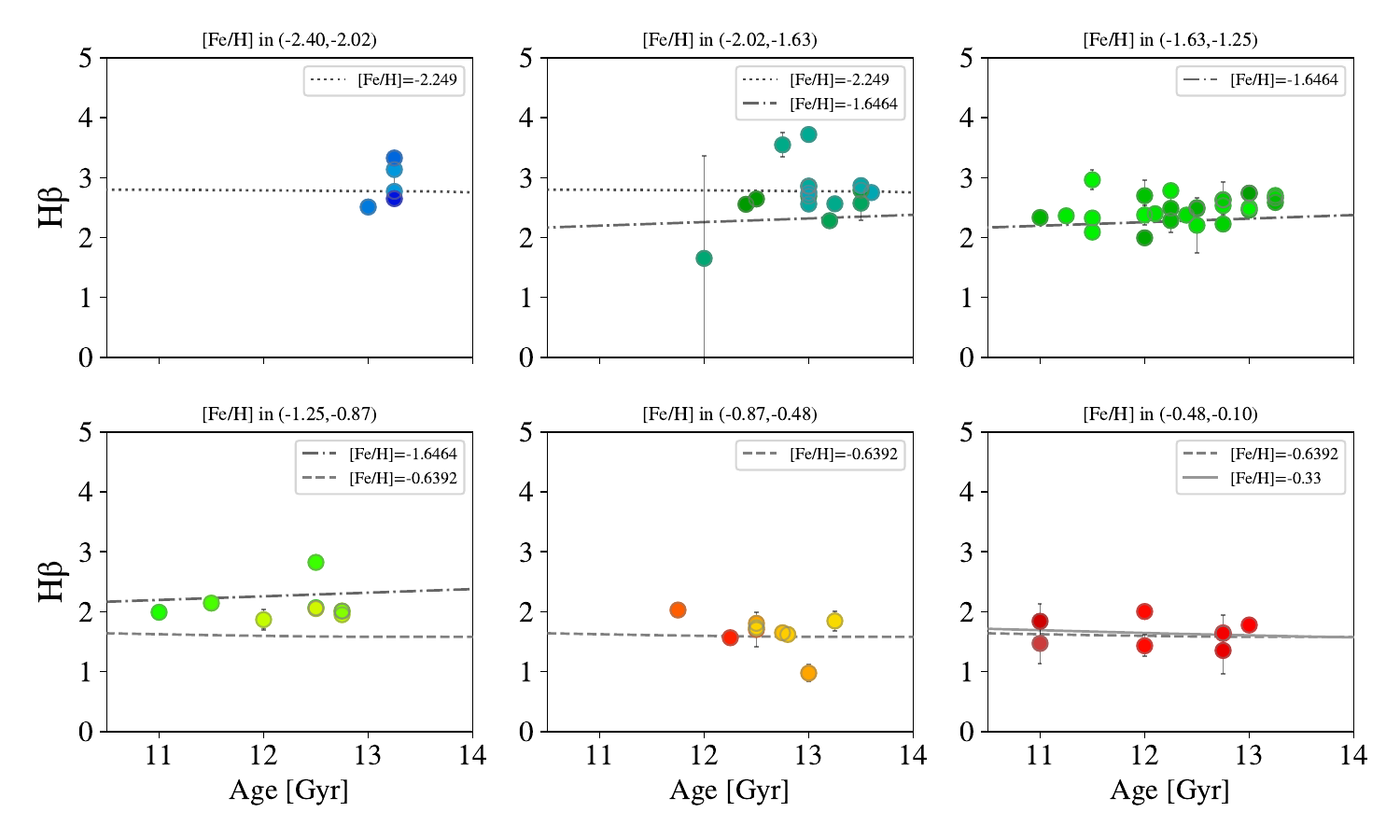}
         \vspace{-0.7cm}
         \caption{}
         \label{fig:Hbeta_age}
    \end{subfigure}
    \caption{$D_n4000$ (top) and H$\beta$ (bottom) trends with age and metallicity. The indices are shown for a sample of 75 GCs for which a literature value of age is available, divided into six [Fe/H] bins and colour-coded according to it. In the background, the stellar models from \citetalias{Bruzual2003} relative to each [Fe/H] bin are shown with different line styles.}
    \label{fig:enter-label}
\end{figure*}

In Fig. \ref{fig:D4000_age} the measured $\mathrm{D_{n}4000}$ is shown as a function of the GC literature age for the 75 GCs for which an age estimate is provided in the literature, divided into six metallicity bins. For a qualitative comparison, we report the theoretical trends from \citetalias{Bruzual2003} with [Fe/H] varying from -0.33 to -2.25 and alpha-enhancement ([$\mathrm{\alpha/Fe}$]) fixed to solar value. The trends are almost flat in this age interval, as expected in these ranges of ages and metallicities \citep[see, e.g.,][]{Moresco2022} but show an evident gradient with metallicity, that is in good agreement with the theoretical distribution, with the $\mathrm{D_{n}4000}$ increasing as metallicity raises. 

Analogous to the $\mathrm{D_{n}4000}$, in Fig. \ref{fig:Hbeta_age} the trends in age are shown for the Balmer index H$\beta$, in comparison with the theoretical trends in the same metallicity bins. The distribution of this feature shows a good agreement with the models as well, this time decreasing with increasing metallicity.

Testing these observables against the stellar models in the case of GCs, objects for which independent and robust measurements of age and metallicities are available, is of great importance in order to validate the models and their use in cosmological analyses. In the application of cosmic chronometers \citep{Jimenez2002}, for instance, especially when the D4000 is directly used to trace the age evolution in redshift \citep{Moresco2012, Moresco2015, Moresco2016} it is fundamental that the D4000 traces correctly the stellar population evolution in time (i.e. the D4000-age slope).  

\begin{figure*}
    \centering
        \begin{subfigure}[b]{\textwidth}
         \centering
         \includegraphics[width=0.45\textwidth]{legendFeH.png} 
    \end{subfigure}
    \begin{subfigure}[b]{0.49\textwidth}
         \centering
         \includegraphics[width=\textwidth]{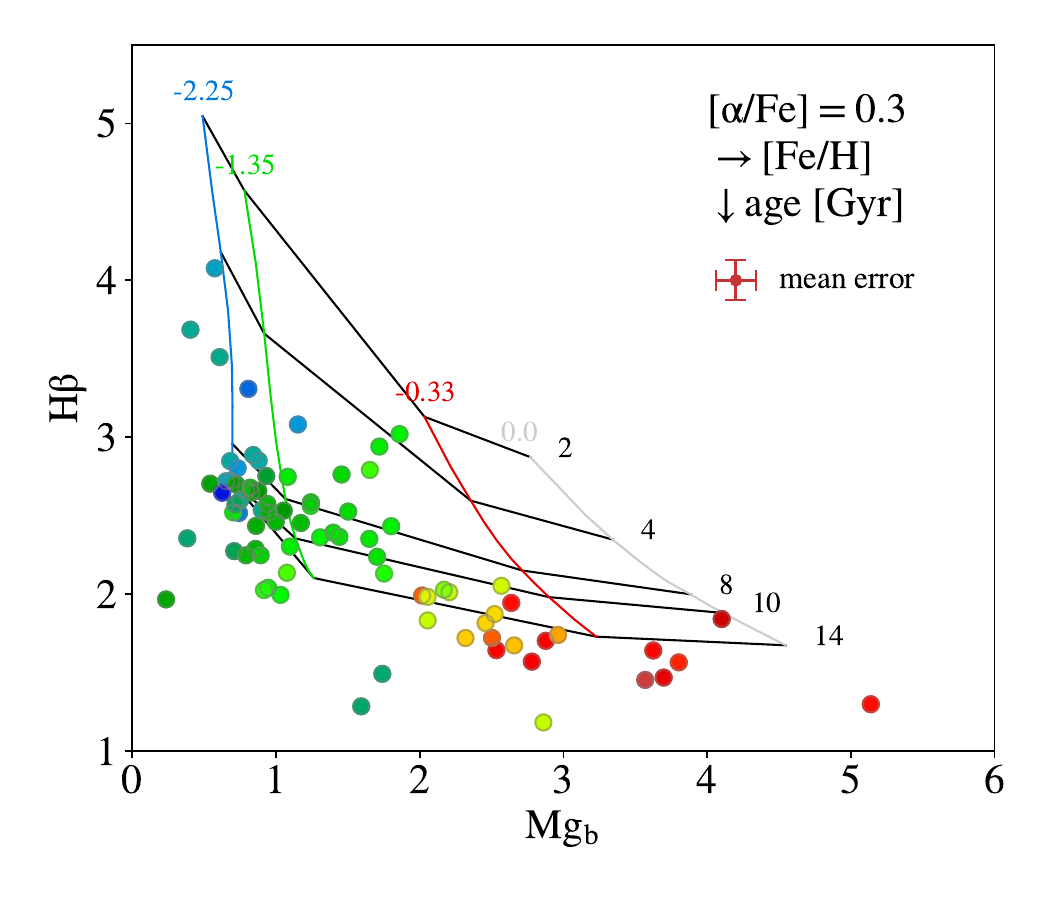}
         \caption{}
         \label{fig:Hb_Mgb}
    \end{subfigure}
    \begin{subfigure}[b]{0.49\textwidth}
         \centering
         \includegraphics[width=\textwidth]{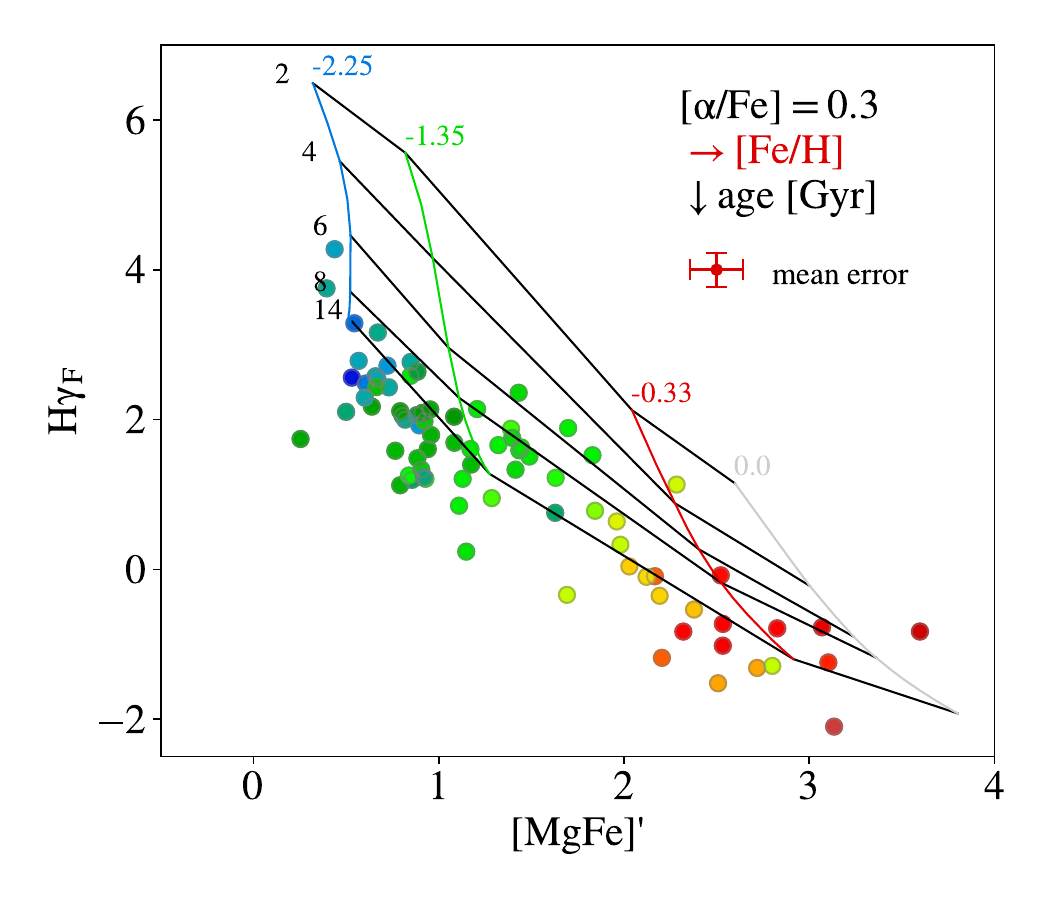}
         \caption{}
         \label{fig:Hg_MgFe}
    \end{subfigure}
    \caption{Index-index diagnostic diagrams, on the left H$\mathrm{\beta}$-$\mathrm{Mg_b}$ and on the right H$\mathrm{\gamma_F}$-[MgFe]'. The points are colour-coded by their literature value of [Fe/H]. We report TMJ models as a grid with varying metallicity (vertical lines) and varying age (horizontal lines), the first with the same colour code as the data points.}\label{Fig:3tmj}
\end{figure*}

\subsection{Index-index diagrams}\label{sec3.2:TMJ_diagrams}
 While in the previous section we studied the sensitivity of single features to the age and metallicity of the population, it is also possible to combine them in the analysis, taking advantage of their different sensitivity to parameters like age, metallicity, and alpha-enhancement. In this case, we are considering a different set of theoretical models to compare them with, specifically developed to take into account also a variation of the chemical composition \citep[][TMJ]{Thomas2011}. Historically, this method has been applied in galaxies \citep[e.g.,][]{Onodera2015, Scott2017, Lonoce2020, Borghi2022a} but also in GCs \citep[e.g.,][]{Strader2004,Proctor2004,Mendel2007,Annibali2018} to derive constraints on the physical properties of these objects. Applying this method, we select and compare indices that are mostly sensible to age, metallicity, or $\alpha$-enhancement variations so that we can disentangle their contributions to the feature's equivalent width. The most widely used are indices of the Balmer series, the iron-group ones, like Fe5270 and Fe5335, and Mgb. As for \citetalias{Bruzual2003}, the models in TMJ have MILES resolution, but differently from \citetalias{Bruzual2003}, they produce a forecast only for the set of Lick indices and not the entire spectrum. In particular, they model the indices values on a grid of ages, metallicities, and alpha-enhancements. Generally, an age-sensitive and a metallicity-sensitive index are plotted one against the other, and compared to a model grid with varying age and [Fe/H] but fixed [$\alpha$/Fe]. In Fig. \ref{Fig:3tmj} we present two examples of these diagrams, H$\mathrm{\beta}$-$\mathrm{Mg_b}$ and H$\mathrm{\gamma_F}$-[MgFe]', where the latter is defined as:
\begin{equation}
    \mathrm{[Mg/Fe]' = \sqrt{\mathrm{Mg_b (0.72 \times Fe5270 + 0.28 \times Fe5335)}}}.
\end{equation}
Here the [$\alpha$/Fe] is fixed at 0.3, which is the closest to the typical value found in the literature for our sample of GCs, around 0.35 \citep[e.g.,][]{Pritzl2005,Mendel2007}. 

This method allows us to have a first estimate of the population's age and metallicity, which aligns well with the literature values, obtained with the independent traditional methods. In terms of age, the sample of GCs populates the area of the oldest objects in the diagrams, with a percentage of data points compatible with an age older than 12 Gyr of 66\% and 88\%, respectively in the H$\mathrm{\beta}$-$\mathrm{Mg_b}$ and H$\mathrm{\gamma_F}$-[MgFe]'. From the diagrams in Fig. \ref{Fig:3tmj} we can also see a small percentage of GCs in the area of typically younger objects, above the 8 Gyr grid-line. In particular, in the H$\mathrm{\beta}$-$\mathrm{Mg_b}$ this happens for 17 GCs, while in the H$\mathrm{\gamma_F}$-[MgFe]' for 11 GCs.

This behaviour can be attributed to the possible presence of an extended horizontal branch (HB), which can make the spectrum appear much bluer than expected for an old population and exhibit prominent Balmer lines, resembling a younger object. This is a very well-known effect, that has always made the study of GCs from integrated light challenging, mostly because the parameters determining the presence and the extent of the HB are not fully predictable with the current stellar evolution models. Various works have made progress in developing diagnostics to identify elongated HBs from integrated light, based on the Balmer lines \citep{Lee2000,Schiavon2004} or on CaII and Mgb \citep{Percival2011}. Others have managed to include the HB contribution on top of the SSP models \citep{Jimenez2004,Koleva2008,CabreraZiri2022}, modelling the emission from the HB hot stars as identified in the GCs' CMD. However, we still lack of a complete modelisation of the HB component, due to the many uncertainties around its origin. Modelling the HB component is beyond the scope of this study, instead, our primary goal is to assess how this unmodeled component may impact studies of integrated populations, using methods commonly employed in galaxy evolution analyses.

\begin{figure}
    \centering
    \includegraphics[width=0.49\textwidth]{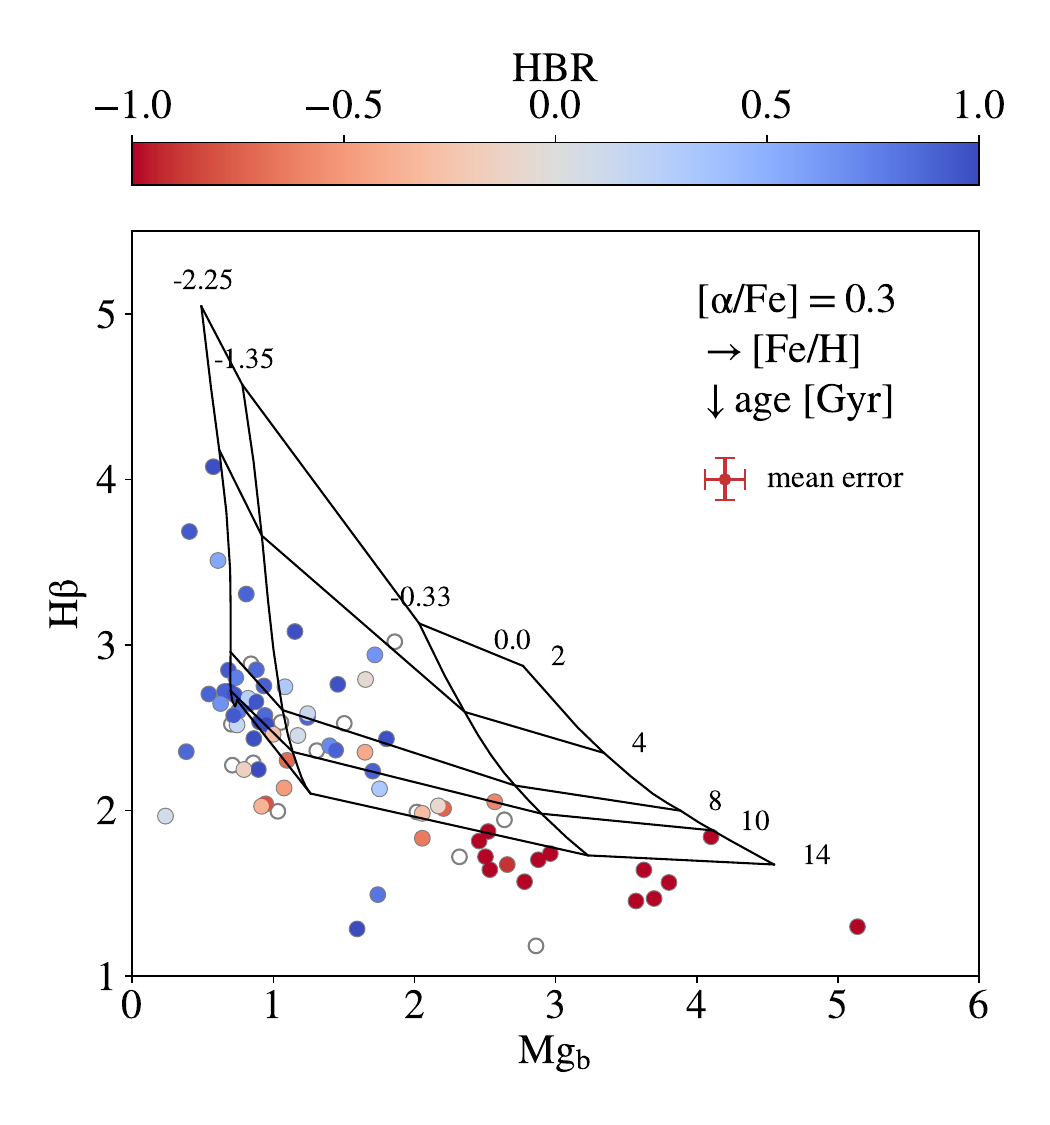}
    \caption{H$\mathrm{\beta}$-$\mathrm{Mg_b}$ diagram, colour-coded by the value of HBR index. Blank points are GCs for which the HBR estimate is not available in \citet{Harris1996} (2010 version). TMJ models are represented as a grid with varying metallicity (vertical lines) and varying age (horizontal lines).}
    \label{fig:3tmj_hbr}
\end{figure}

The most common parameter used to quantify the HB extent is the morphology index HBR \citep{Lee1989,Lee1994}, defined as:
\begin{equation}
    \mathrm{HBR = \frac{B-R}{B+R+V}}
\end{equation}
where B and R are the number of stars bluer and redder than the RR Lyrae instability strip, and V is the number of RR Lyrae stars. Although this parameter does not fully capture the distribution of stars along the HB, it still provides valuable information about the HB morphology, indicating whether it is predominantly red (HBR$\sim-$1) or blue (HBR$\sim$1). In Fig. \ref{fig:3tmj_hbr} we report the same H$\mathrm{\beta}$-$\mathrm{Mg_b}$ diagram as in Fig. \ref{fig:Hb_Mgb}, but this time colour-coded by the HBR value listed in \citet{Harris1996} (2010 edition), which is known for 69/82 objects (coloured points). According to this index, among the 14 GCs populating the area above the 8 Gyr grid line and for which the HBR is known, 13 show a blue HB, and for 11 of those HBR is even higher than 0.5, a clue of a very elongated blue HB. The fraction of objects with HBR above zero decreases as we move to the areas belonging to older ages: 72\% (18/25) between 8 Gyr and 12 Gyr and 33\% (10/30) over 12 Gyr. The same trend can be observed moving from lower to higher [Fe/H], with a percentage of GCs showing blue HBs decreasing from 68\% (23/34) at $\mathrm{[Fe/H]<-1.35}$ to 49\% (13/29) in the range -1.35 $\mathrm{\leq[Fe/H]\leq}$ -0.33 and dropping to zero at $\mathrm{[Fe/H]>-0.33}$. 

As anticipated, stellar evolution models do not currently account for the presence of an extended blue HB, so this must be considered in the FSF analysis, where objects with extended HB morphology might be mistakenly identified as young stellar populations \citep{Schiavon2004,Jimenez2004,Koleva2008,CabreraZiri2022}. From our initial qualitative analysis using indices, we expect this issue to be more prevalent in metal-poor objects with prominent \(\mathrm{H\beta}\), which tend to show the highest HBR. 

%%%%%%%%%%%%%%%  METHOD #########################

\section{Method and analysis}
\label{sec:4METHOD}
In this section, we present the method adopted to estimate the ages and physical properties of the GCs sample, the code used, its settings, and the results obtained.

\subsection{Full-spectral-fitting with \texttt{BAGPIPES}}
\label{sec:3.1FSFwithBagpipes}
We perform FSF using the public code \texttt{BAGPIPES} \citep{Carnall2018}, which allows us to fit spectra and/or photometry adopting a parametric Bayesian approach. A detailed description of all the code's features is presented in \citet{Carnall2019} and \citet{Carnall2022}, while an overview of the settings that we adopt is already outlined in \citet{Tomasetti2023}. Here we recap the main features of the code, highlighting the aspects that are relevant to this work. 

\texttt{BAGPIPES} is able to model synthetic spectra and photometry, based on a set of instructions, and then fit the so-modelled spectro-photometry to the observed one via a Bayesian approach, thus maximising the posterior probability using a nested sampling algorithm, \texttt{Multinest} \citep{Buchner2016}. In this work, we focus on four main model components to construct the synthetic spectra.

The first component is an SSP model, which is designed to reproduce the continuum emission and the absorption features of a population built up in a single episode of star formation. The SSP models implemented in \texttt{BAGPIPES} are the 2016 version of \citet{Bruzual2003} \citep[BC16, see][]{Chevallard2016}. They produce different synthetic spectra based on the wavelength $\lambda$ range, the age of the stellar population, and its overall metallicity [Z/H], assuming a \citet{Kroupa2001} initial mass function (IMF). 
    
The second component is the star formation history (SFH). The code allows you to combine different SFHs, one for each SSP, but dealing with GCs we implement a single SFH, assuming a unique star formation episode. In particular, we adopt the delayed exponentially declining (DED) SFH, which is given by the equation
\begin{equation}\label{eq:3delayedSFH}
        \mathrm{SFR(t) \propto} \begin{cases} 
        \mathrm{(t-T_0)}\: \mathrm{e^{-\frac{t-T_0}{\tau}}}, & \mathrm{t > T_0} \\ 0, & \mathrm{t < T_0} 
        \end{cases},
\end{equation}
where SFR is the star formation rate, $\tau$ provides the width of the SFH and $\mathrm{T_0}$ sets the age of the Universe at which the star formation begins. Using a single DED is recommendable when dealing with a stellar population whose time scale of formation is much shorter than its age, as we expect for GCs. 
    
The third component is dust absorption and emission. This is particularly important to model the redder part of the spectrum, which can be largely depressed due to the presence of dust in the system. In the context of MW GCs, this component is necessary to account for the MW dust on the line-of-sight, for which WAGGS spectra are not corrected. The model implemented here is the \citet{Salim2018}, represented by a power-law, as in \citet{Calzetti2000}, with an additional parameter, $\delta$, representing a slope deviation.

The last component is a non-physical term representing noise, which can be added to the error spectrum to account for any potential underestimation. This noise is introduced as white noise.

After running the code, we obtain a best-fit spectrum and the posterior distributions for all the parameters involved, like age, mass formed, overall metallicity, and dust extinction. At the same time, \texttt{BAGPIPES} can provide estimates of derived quantities, like the SFR or the stellar mass formed, parameters that are not directly involved in the fit. In particular, the mass formed ($\mathrm{M_{formed}}$) and the stellar mass formed ($\mathrm{M_{\star}}$) are different in that the first one comprises all the mass formed from $t=0$ to the time $t$ at which the GC is observed:
\begin{equation}
    \mathrm{M_{formed} = \int_0^t SFR(t')dt'},
\end{equation}
including also stellar remnants ($\mathrm{M_{rem}}$), while the second one includes only the mass of living stars: 
\begin{equation}
    \mathrm{M_{\star} = \int_0^t SFR(t')dt' - M_{rem}}.
\end{equation}
From now on, we refer to $\mathrm{M_{\star}}$ as the mass of the objects.

Since we aim to use the resulting ages in a cosmological framework, it is important to avoid any constraint based on cosmology. For this reason, we employ a version of \texttt{BAGPIPES}, described in \citet{Jiao2023}, that deviates from the original in handling the priors on the stellar population age, allowing them to vary up to a cosmology-independent value (e.g., 15 Gyr, 20 Gyr) at any redshift. This modification has already been tested and validated in VANDELS \citep{Tomasetti2023} and LEGA-C \citep{Jiao2023}. Originally the code assumes a cosmological prior on ages, for which the maximum age resulting from the fit should be smaller than the age of the Universe at the corresponding redshift, given a flat $\Lambda$CDM model with parameters $\mathrm{\Omega_{M}=0.3,\: \Omega_\Lambda=0.7\: and\: H_0=70\: km\:s^{-1}Mpc^{-1}}$. Although this effect is of relative interest in stellar population studies and is typically neglected, it can't be ignored in cosmological analyses because the derived ages would be constrained by the cosmological model assumed, leading to results that just recover the assumed prior. Below, we are going to test the effect of different assumptions of age prior on the sample of GCs.   

\subsection{Full-spectral-fitting in WAGGS}
\label{sec:fsf_waggs}
Before inputting the cluster spectra into BAGPIPES, some adjustments were necessary. 

First, we downgraded the spectral resolution to approximately 2.7 {\AA} FWHM, consistent with the \citetalias{Bruzual2003} models used in the code.

Next, we aligned the spectra to the correct frame using distances from \citet{Baumgardt2023} and corrected for radial velocity variations, which could cause minor blueshifts or redshifts in the spectra.

To prevent underweighting the blue features in the fit — due to the non-uniform error spectrum, with S/N ranging from a few tens to a thousand — we set an upper limit for the S/N at 100 and adjusted the error spectrum accordingly. We tested various S/N thresholds (e.g., 20, 50), finding that they had minimal impact on the results, except when the S/N ratio between the blue and red ends of the spectrum differed by more than a factor of ten, which resulted in very low weight for the blue features in the fit.

After these adjustments, we performed multiple tests to optimally use \texttt{BAGPIPES} on GCs spectra and find the best-fit configuration to reproduce their spectral features accurately. In particular, this process involved: adopting different SFHs (e.g., single burst, delayed exponentially declined) with different priors on the parameters; fitting different wavelength ranges, either moving the lower limit to longer wavelengths to reduce the contamination by HB stars or pushing the upper limit to redder features to better constrain dust reddening; testing different priors on the GC's [Z/H] and mass (e.g., uniform, Gaussian, logarithmic) to assess their impact on the estimation of these parameters, as well as the influence on ages, given the degeneracies at play. It is worth mentioning here that the mass and metallicity estimates proved to be very stable against all the different changes in the fit setup, while ages were mainly affected by the choice of prior, as we discuss in the following. \\In the end, we converged to a fit configuration in which: \textit{(i)} as often done in literature \citep{Koleva2008,Goncalves2020} we fit the range 3700-6000 $\AA$ to avoid the redder telluric lines and we mask the interval $\mathrm{5870 - 5910}$ {\AA}, where the spectra show a very deep sodium doublet absorption line, since it could be potentially contaminated by interstellar absorption; \textit{(ii)} we consider a single DED SFH, a dust component and a noise component; \textit{(iii)} on all the parameters we set uniform, uninformative priors.

To assess the impact of the cosmological prior on the results, we tested two different upper limits for the age parameter: 13.47 Gyr, age of the Universe in a flat $\Lambda$CDM model with $\mathrm{\Omega_{M,0}}$ = 0.3, $\mathrm{\Omega_{\Lambda,0}}$ = 0.7, $\mathrm{H_0}$ = 70 km/s/Mpc; 15 Gyr, a loose limit independent of any cosmology. We refer to the first configuration as {\it Config. 13.5} and to the latter as {\it Config. 15}. A summary of the main parameters and relative priors for the two configurations can be found in Tab. \ref{Tab:3priors}. We highlight that we assume uniform priors on all the parameters, along with wide ranges so that the results are not constrained by any previous knowledge of the GC's mass, metallicity, dust extinction or age.

As anticipated, BAGPIPES adopts the overall metallicity [Z/H] as the metallicity parameter. To compare our results with [Fe/H] values from the literature, we need to perform a conversion. We use the conversion formula from \citet{Salaris2005}:
\begin{equation}\begin{split}
    \mathrm{[Z/H]\:} &\mathrm{= log\left(\frac{Z}{Z_\odot}\right) =} \\ &\mathrm{\:= [Fe/H] + log_{10}\left(10^{[\alpha/Fe]} 0.694 + 0.306\right)}.
\end{split}
\end{equation}
For objects with $\mathrm{[Fe/H] \leq-1}$ we apply this formula with an alpha-enhancement of [$\alpha$/Fe]=0.35, which is typical of the metal-poor MW GCs \citep[e.g.,][]{Pritzl2005,Mendel2007}, while for GCs with $\mathrm{[Fe/H]>-1}$ we use [$\alpha$/Fe]=0.15, average alpha-enhancement at these metallicities \citep[see, e.g.,][]{Pagel1995,Pancino2017}. From now on, we refer to the quantity [Z/H] as the metallicity of the GCs.

\begin{table*}[]
\centering
\caption{Parameters and priors for different configurations.}\label{Tab:3priors}
\begin{tabular}{@{}ccccccccc@{}}
\hline\midrule
             & $\lambda$ range [{\AA}] & age [Gyr] & SFH & $\tau$ [Gyr] & $\mathrm{log(M_{\star}/M_{\odot})}$ & [Z/H]        & dust $\mathrm{A_V [mag]}$ & sigma [km/s] \\\midrule
Config. 13.5 & 3700 - 6000            & U(0,13.8) & DED & U(0,0.15) & U(1,6.3)     & U(-2.3,0)    & U(0,4)   & U(0,15)      \\
Config. 15   & 3700 - 6000            & U(0,15)   & DED & U(0,0.15) & U(1,6.3)     & U(-2.3,0)    & U(0,4)   & U(0,15)      \\
1 & 3700 - 6000 & U(0,15) & DED   & U(0,1)    & U(1,6.3) & U(-2.3,0) & U(0,4) & U(0,15) \\
2 & 3700 - 6000 & U(0,15) & burst & -         & U(1,6.3) & U(-2.3,0) & U(0,4) & U(0,15) \\
3 & 3350 - 9000 & U(0,20) & DED   & U(0,0.15) & U(1,6.3) & U(-2.3,0) & U(0,4) & U(0,15) \\
4            & 3700 - 6000            & U(0,15)   & DED & U(0,0.15) & U(1,6.3)     & G(val, 0.01) & U(0,4)   & U(0,15)      \\
5 & 3700 - 9000 & U(0,15) & DED   & U(0,0.15) & U(1,6.3) & U(-2.3,0) & U(0,4) & U(0,15) \\
6 & 3700 - 6800 & U(0,15) & DED   & U(0,0.15) & U(1,6.3) & U(-2.3,0) & U(0,4) & U(0,15) \\
7 & 4828 - 5634 & U(0,15) & DED   & U(0,0.15) & U(1,6.3) & U(-2.3,0) & U(0,4) & U(0,15) \\
8 & 3700 - 6000 & U(0,15) & DED   & U(0,0.15) & -        & U(-2.3,0) & U(0,4) & U(0,15) \\ \bottomrule
\end{tabular}
\end{table*}

\subsection{Results}\label{sec:4results}
We performed a visual inspection to evaluate the quality and convergence of the fits. Specifically, we identified fits that either failed to recover the spectral lines or continuum or exhibited double- or multiple-peaked posterior distributions. As a result, we discarded a significant number of poor fits, totalling 11 objects in both {\it Config. 13.5} and {\it Config. 15}, which represent about 14\% of the sample. Among these, 8 GCs had an HBR$>$0, and we found that, in these cases, the posterior spectrum underestimated the emission in the wavelength range blueward of $\mathrm{4000-4500\: \AA}$. This issue is likely due to the blue HB emission, which the models cannot fully reproduce. Consequently, the fits converge to younger ages, as observed in these cases where all 8 poor fits have ages younger than 10 Gyr.  As discussed in Sect. \ref{sec3.2:TMJ_diagrams}, various studies have successfully included a contribution of the HB on top of the SSP models \citep[e.g.,][]{Jimenez2004,Koleva2008,CabreraZiri2022}. However, incorporating this component into \texttt{BAGPIPES} is outside the scope of this work. Removing bad fits, the clean sample counts 66 GCs in both configurations. 

\begin{figure*}
    \includegraphics[width=\textwidth]{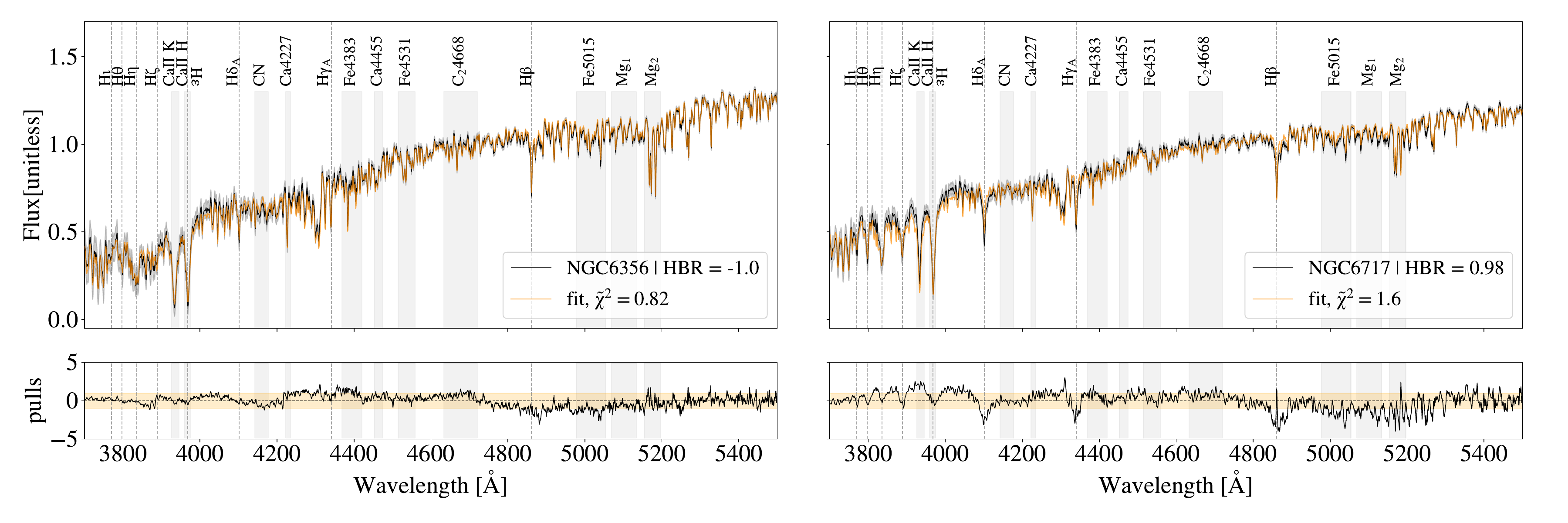}
    \caption{Examples of good fits. In the top panels, the observed spectra are shown in black and the posterior ones in orange, dashed lines identify the Balmer absorption series, while other main absorption features are highlighted with grey shaded boxes. In the bottom panels, the pulls of each fit ((observed - fit)/error) are shown, with the orange horizontal area representing a 1-$\sigma$ fluctuation.}\label{fig:4fit_examples}
\end{figure*}

In Fig. \ref{fig:4fit_examples} two examples of good fits are reported, both converging to ages older than 13 Gyr, one presenting a red HB (NGC6356, HBR=-1) while the other shows a blue HB (NGC6717, HBR=0.98). The pulls highlight how in the case of NGC6356 the stellar models, plus the dust components, are able to accurately reproduce the GC's spectrum, with residuals compatible with 1-$\sigma$ fluctuations at all wavelengths. In the case of NGC6717, the quality of the fit is still good, but the pulls clearly show a residual at bluer wavelengths, especially concerning the Balmer absorption lines, pointing out the unmodelled hot stars component. This suggests that for GCs characterised by blue HBs may still produce a reliable age estimation, as long as the blue HB stars do not outshine the blue end of the spectrum.

The quality of both considered setups is highlighted by the median reduced chi-squares, $\mathrm{\Tilde{\chi}^2=1.21}$ in {\it Config. 13.5} and $\mathrm{\Tilde{\chi}^2=1.26}$ in {\it Config. 15}. This is quite noticeable since in this case we adopted the formal spectrum error provided by the analysis, with the correction described in Sect.~\ref{sec:fsf_waggs}. These values are further (and as expected) reduced if we take into account the noise parameter, which acts in correcting the error spectrum for potential underestimations, leading to $\mathrm{\Tilde{\chi}^2=0.99}$ in {\it Config. 13.5} and $\mathrm{\Tilde{\chi}^2=0.98}$ in {\it Config. 15}.

To analyse the derived physical properties, we consider, for each parameter, the median and the $\mathrm{16^{th}}$ and $\mathrm{84^{th}}$ percentiles of the posterior distribution respectively as the best-fit value, lower and upper error. 

\begin{figure*}
    \centering
    \begin{subfigure}[b]{\textwidth}
         \centering
         \includegraphics[width=0.45\textwidth]{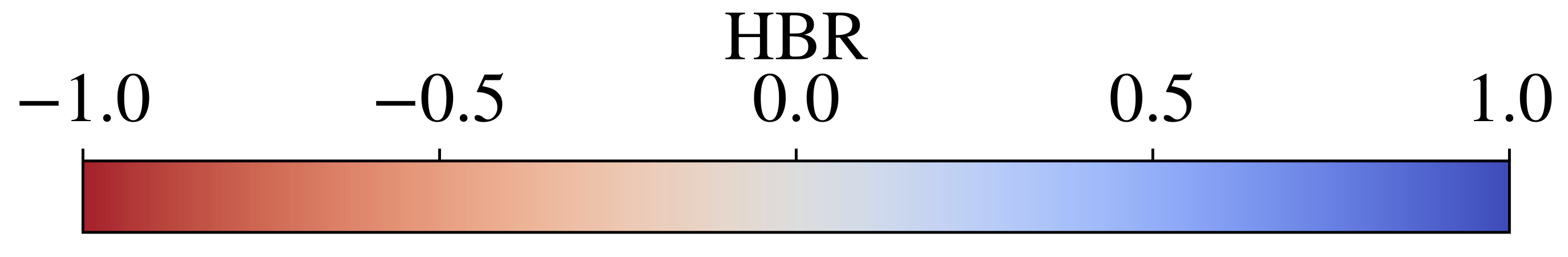} 
    \end{subfigure}
    \begin{subfigure}[b]{0.49\textwidth}
         \centering
         \includegraphics[width=\textwidth]{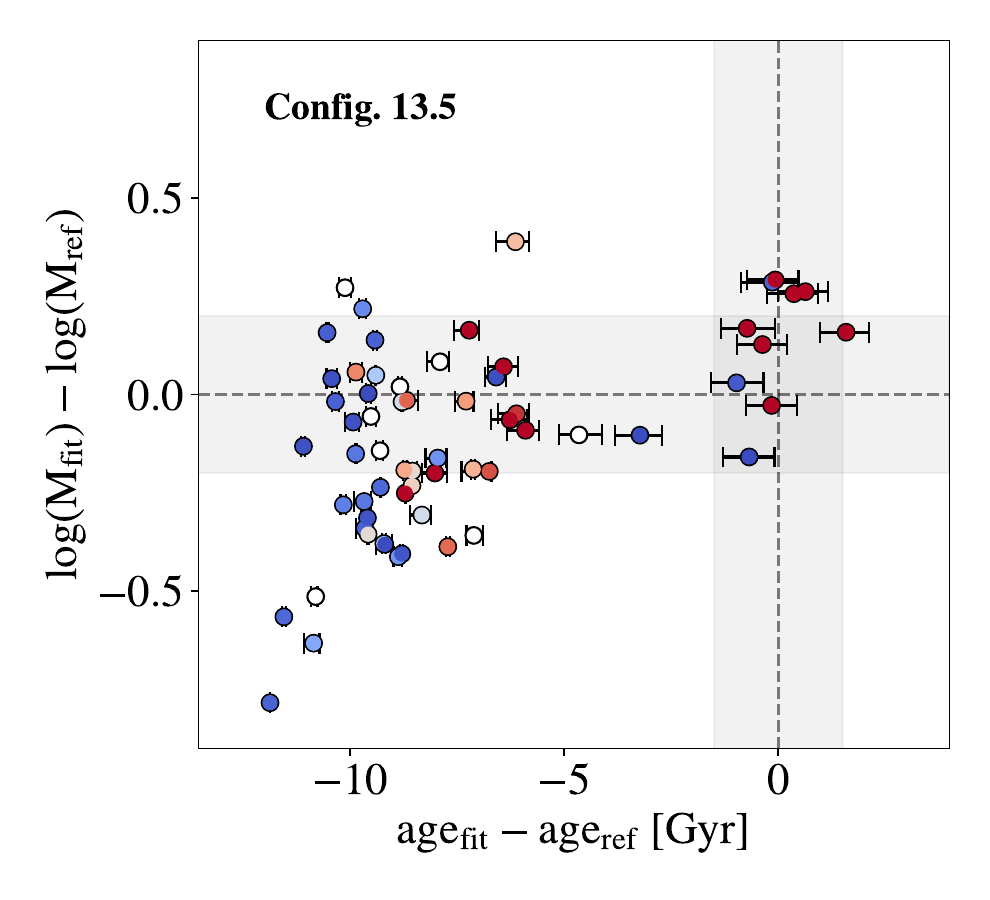}
         \caption{}
         \label{fig:mass_age_13.5}
    \end{subfigure}
    \begin{subfigure}[b]{0.49\textwidth}
         \centering
         \includegraphics[width=\textwidth]{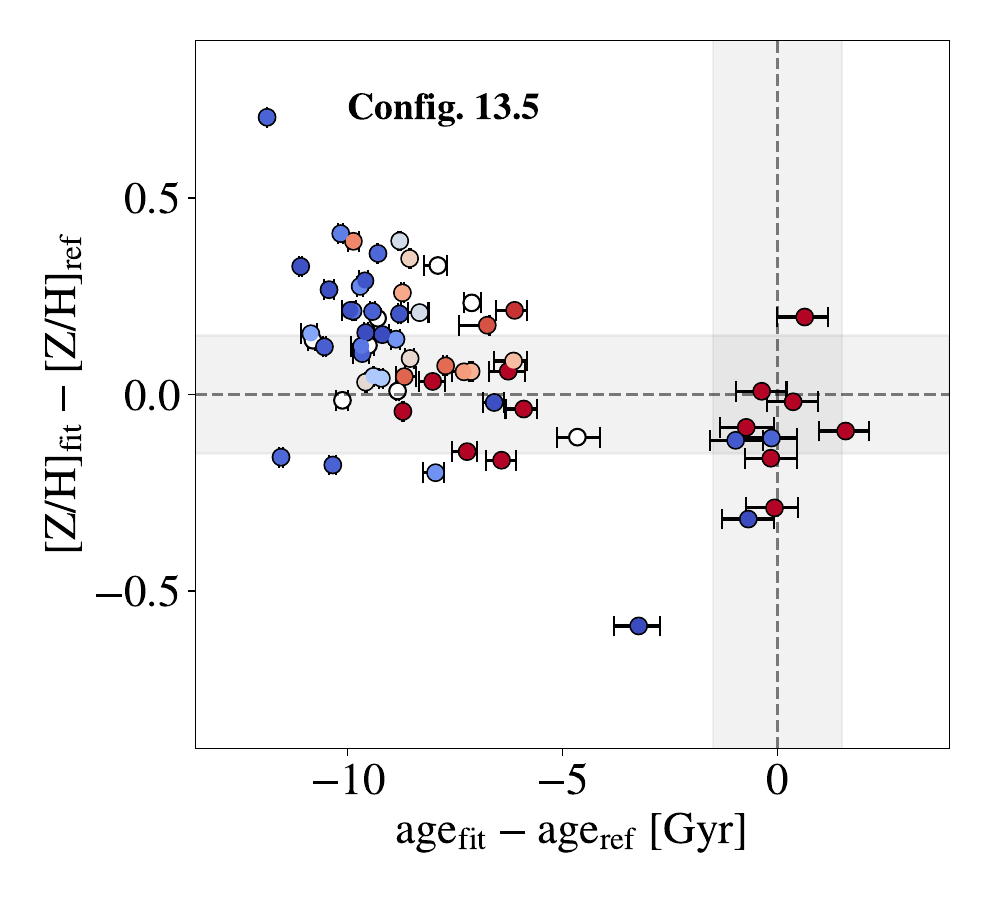}
         \caption{}
         \label{fig:met_age_13.5}
    \end{subfigure}
    \begin{subfigure}[b]{0.49\textwidth}
         \centering
         \includegraphics[width=\textwidth]{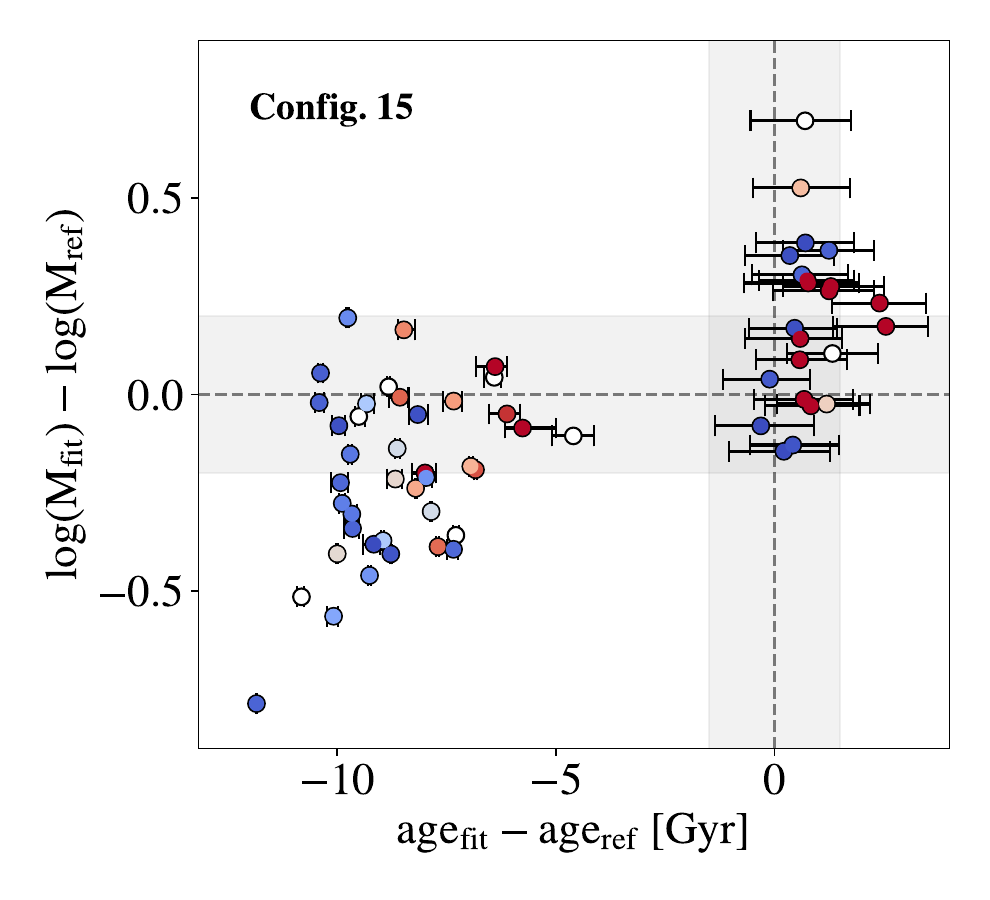}
         \caption{}
         \label{fig:mass_age_15}
    \end{subfigure}
    \begin{subfigure}[b]{0.49\textwidth}
         \centering
         \includegraphics[width=\textwidth]{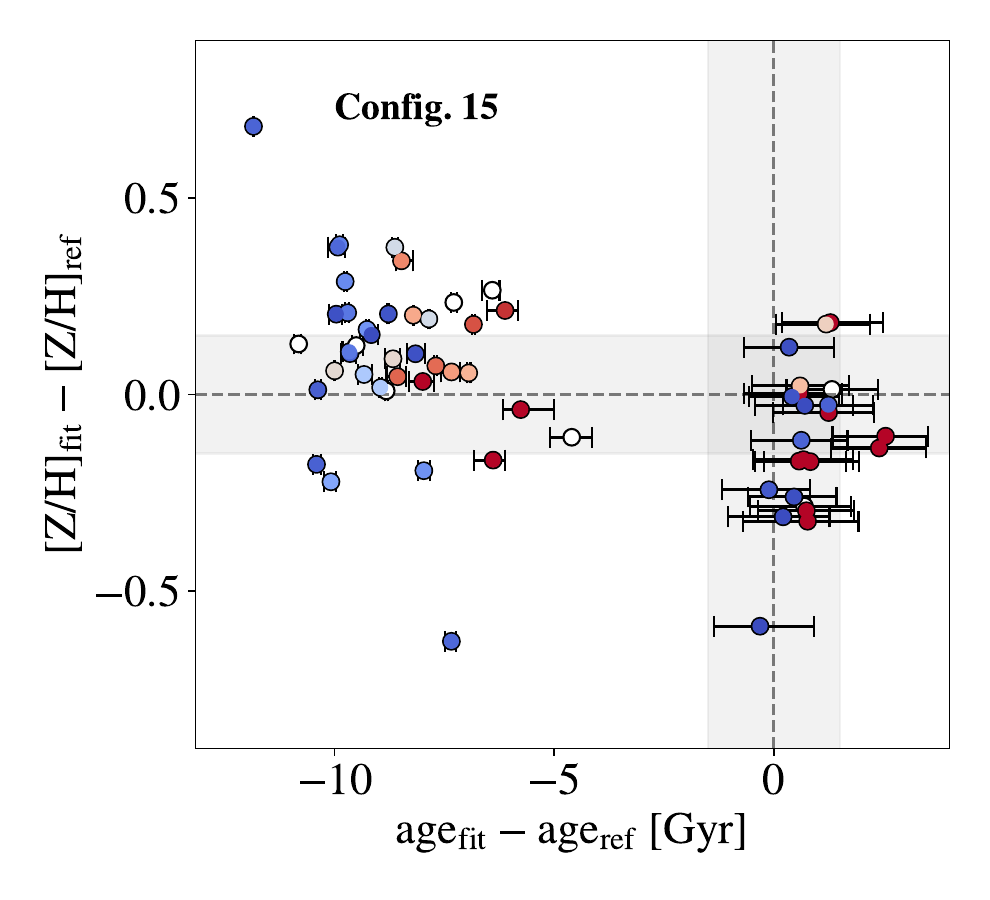}
         \caption{}
         \label{fig:met_age_15}
    \end{subfigure}
    \caption{Differences in stellar mass ($\mathrm{\Delta \mass}$) and metallicity ($\mathrm{\Delta [Z/H]}$) as a function of the difference in age as estimated in this work with respect to literature values. The dashed lines correspond to a null difference, the grey shaded areas represent an average representative error on literature values, namely 0.2 dex in mass, 0.15 dex in metallicity and 1.5 Gyr in age. The top two panels refer to {\it Config. 13.5} while the bottom ones to {\it Config. 15}. All the points are colour-coded by their HBR index.}\label{Fig:4results}
\end{figure*}

\subsubsection{Configuration 13.5}
In {\it Config. 13.5} we observe that metallicities and GC masses are in good agreement with literature values, with mean deviations of $\mathrm{\langle \Delta [Z/H]  \rangle = 0.09 \pm 0.21}$ dex and $\mathrm{\langle \Delta \mass \rangle = -0.09 \pm 0.24}$ dex respectively, consistent with the typical errors associated with these quantities (see Sect. \ref{sec:2DATA}). In terms of stellar age, instead, a clear bimodality is present. While 17\% of the sample (11 GCs) turns out to have ages older than 10 Gyr and only $\sim$0.16 Gyr younger than literature values on average, most of it (55 GCs) shows ages significantly younger than 10 Gyr, $\sim$8.9 Gyr lower on average. We investigate this difference in the following.

In Fig. \ref{fig:mass_age_13.5} and \ref{fig:met_age_13.5} we show the differences in stellar mass and metallicity as a function of this age gap, colour-coded by HBR index. This highlights two important aspects. \\The first is, again, the trend with HBR, showing that when this index is positive (blue HB), the code misinterprets the blue shape of the spectrum and the deep Balmer lines for a young population 87\% of the time (27/31), resulting in ages on average 8.4 Gyr younger than expected from the literature. This exact behaviour is also observed for most of the red HB population, but in a smaller fraction (73\% of the cases, 19/26 GCs) and with a less significant age discrepancy of 5.1 Gyr on average. A similar result was already found both in \citet{Koleva2008} and \citet{CabreraZiri2022}, where the issue was mitigated by adding a fraction of hot stars on top of the SPS models, but, as anticipated, including the HB component goes beyond the purpose of this work. In Sect. \ref{sec:CC22comparison}, though, a detailed comparison with the results in \citet{CabreraZiri2022} can be found. \\The second is the existence of a degeneracy among the parameters involved. A lower cluster mass or a higher metallicity can easily mislead the fit to ages much younger than the literature one. This is clear if we compute the median differences in metallicity and mass separately for the GCs resulting older and younger than 10 Gyr: for the former, we find average deviations of $\mathrm{\langle \Delta [Z/H] \rangle = -0.14 \pm 0.19}$ dex and $\mathrm{\langle \Delta \mass \rangle = 0.16 \pm 0.15}$ dex, for the latter instead $\mathrm{\langle \Delta [Z/H] \rangle = 0.13 \pm 0.18}$ dex and $\mathrm{\langle \Delta \mass \rangle = -0.13 \pm 0.24}$ dex.

\subsubsection{Configuration 15}
In {\it Config. 15} the good agreement of metallicity and GC mass estimates with literature values holds, with average differences of $\mathrm{\langle \Delta [Z/H] \rangle = -0.02 \pm 0.23}$ dex and $\mathrm{\langle \Delta \mass \rangle = 0.04 \pm 0.28}$. Concerning the stellar ages, even though the only difference with respect to {\it Config. 13.5} is the removal of the cosmological prior, the results improve significantly. Here 36\% (24 GCs) of the sample shows ages older than 10 Gyr, more than twice the old population of {\it Config. 13.5}. Among these 24 GCs, the ages result compatible with literature values 92\% of the times (22 GCs), on average 0.67 Gyr older, and we find average discrepancies in [Z/H] and mass of $\mathrm{\Delta[Z/H]=-0.11 \pm 0.18}$ and $\mathrm{\Delta \mass=0.20 \pm 0.20}$. In Fig. \ref{fig:mass_age_15} and Fig. \ref{fig:met_age_15} we show the analogous of \ref{fig:mass_age_13.5} and Fig. \ref{fig:met_age_13.5} for {\it Config. 15}. We can see that both the trend in HBR and the age-metallicity and age-mass degeneracies are present but with important differences. This time, removing the upper limit on the age parameter has reduced the fraction of blue HB GCs mistaken for young populations to 71\% (22/31) and the one of the red HB GCs to 54\% (14/26). This means that 13 GCs previously resulting younger than 10 Gyr are now recognised as old populations, representing a 20\% increment. The degeneracies cited above play an important role in this because all of these 13 GCs are here characterised by a lower metallicity ($\mathrm{\Delta[Z/H] \sim - 0.18\: dex}$) and a higher mass ($\mathrm{\Delta \mass \sim 0.19\: dex}$) than the one found in {\it Config. 13.5}, yet mostly in agreement with literature values within errors. This suggests that for a fraction of GCs an old, more realistic solution does exist beyond the cosmological limit usually set when performing FSF and that it may also be preferred to the younger one if this area of the parameter space is made accessible. For this reason and in light of the subsequent cosmological analysis, we consider {\it Config. 15} our benchmark.

\subsection{Systematic effects}\label{sec:4syst}
As anticipated in Sect. \ref{sec:fsf_waggs}, we performed multiple tests with different settings to find the optimal fit configuration. These analyses have been used to assess the systematic error induced in the age determination with our approach. Specifically, we examined 8 configurations, each differing from our benchmark in one or two key aspects, including variations in SFH type (burst or DED), age prior (15 Gyr or 20 Gyr) metallicity prior (uniform or Gaussian), wavelength range of the fit, and in fitting spectra in physical units or normalised in the window 4500-5000 {\AA}. The latter was applied in just one configuration, the only case in which the mass parameter could not be determined due to the normalisation. All the characteristics of the different configurations are outlined in Tab. \ref{Tab:3priors}, numbered from 1 to 8. 

In this analysis, we discarded all the spurious young solutions with a best-fit age below 10 Gyr for the same reasons discussed in Sect. \ref{sec:4results}, and all the bad fits in \textit{Config. 15} and in each of the 8 test configurations. In this way, we end up with a sample of 18 GSs having a good fit in at least 5 out of the 9 runs. For each object, we computed the standard deviation of the age distribution in the 9 runs. Finally, we estimated a global systematic contribution to the age uncertainty as the average of these standard deviations, weighted on the number of good fits for each GC, resulting in 0.71 Gyr\footnote{We note here that taking the median provides a similar, yet a bit more optimistic result, estimating a 0.64 Gyr systematic contribution to the error budget.}.

\subsection{The role of metallicity}
\begin{figure}
    \centering
    \begin{subfigure}[b]{0.49\textwidth}
         \centering
         \includegraphics[width=\textwidth]{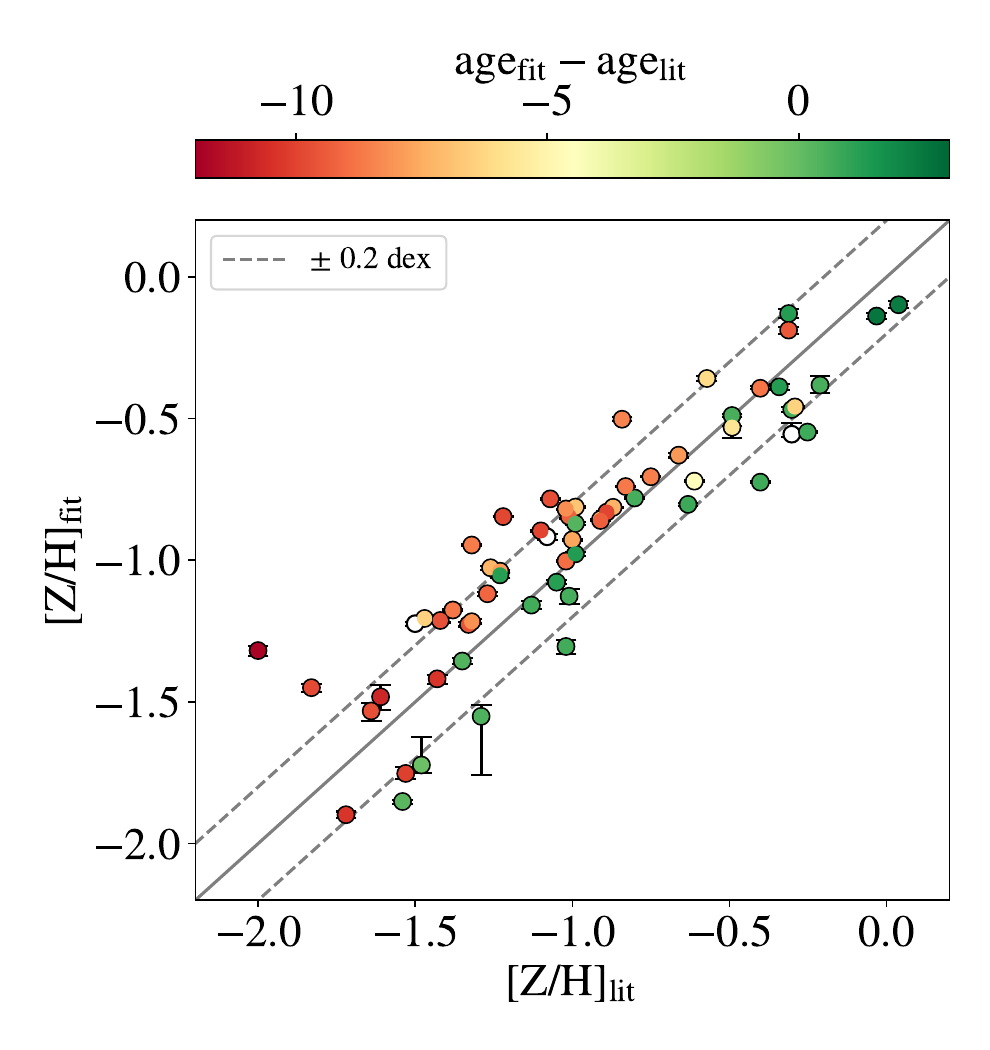}
         \caption{}
         \label{fig:met_met}
    \end{subfigure}
    \begin{subfigure}[b]{0.47\textwidth}
         \centering
         \includegraphics[width=\textwidth]{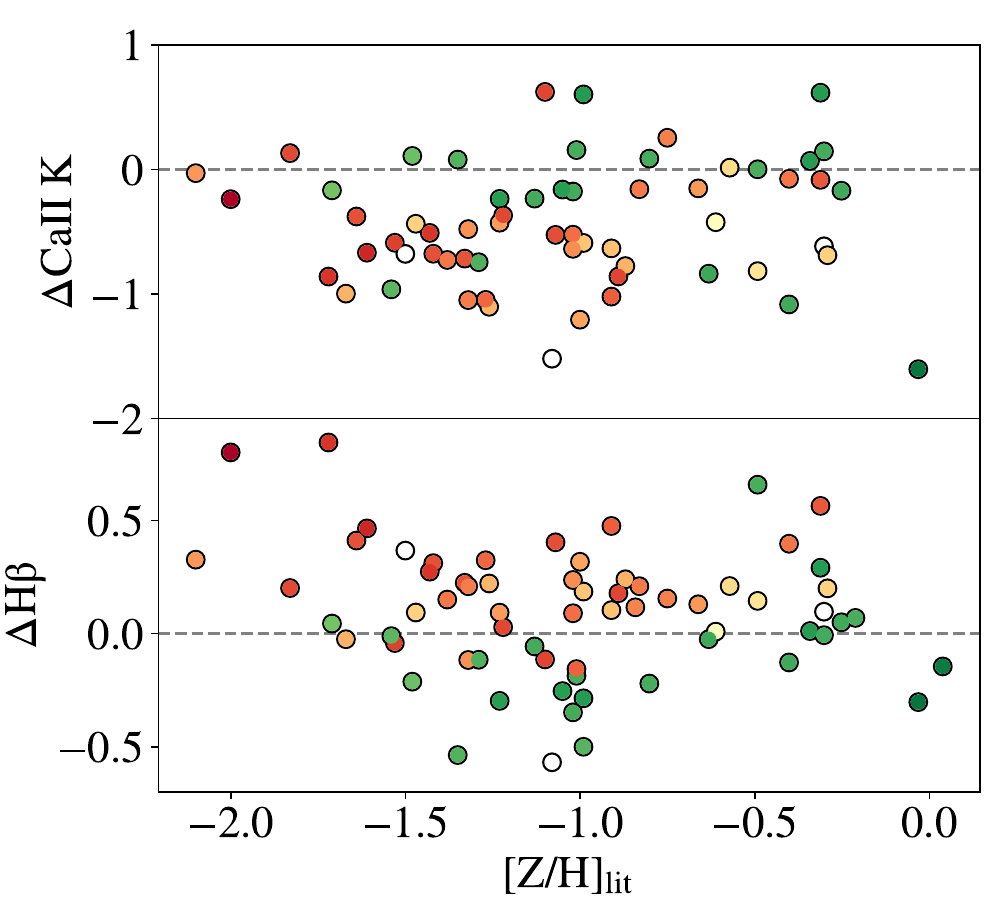}
         \caption{}
         \label{fig:CaIIK_Hbeta}
    \end{subfigure}
    \caption{Metallicity and differences in CaII K and H$\beta$ measurements as a function of literature metallicity. \textit{Top}: [Z/H] obtained in this work against literature [Z/H], colour-coded by the corresponding difference in age. The continuous line represents the one-to-one relation, and the dashed lines the 0.2 dex scatter. \textit{Bottom}: Differences in CaII K and H$\beta$ as measured on the posterior spectra with the observed ones, colour-coded by the difference in age.}
\end{figure}

In Sect. \ref{sec3.2:TMJ_diagrams} we observed that not only the presence of an extended blue HB, but also a low metallicity could produce some spectral features that can drag the fit to younger ages due to their degeneracy. Here we want to verify if this trend is present in our results. In Fig. \ref{fig:met_met}, we plot the metallicity obtained from our fit against the values found in the literature, colour-coded by the difference between the age estimated from the fit and the literature one. We can observe that the best match in age estimation is indeed found at the highest metallicities, while the objects for which the age is most underestimated are also the ones with lower metallicity. To perform a quantitative comparison, we divided our sample into metallicity sub-samples, considering three intervals equally spaced in $\mathrm{[Z/H]_{lit}}$: a metal-rich $\mathrm{[-0.7,0.0]}$, a metal-intermediate $\mathrm{[-1.4,-0.7]}$ and a metal-poor $\mathrm{[-2.1,-1.4]}$. 

\begin{figure*}
    \centering
    \begin{subfigure}[b]{\textwidth}
         \centering
         \includegraphics[width=0.95\textwidth]{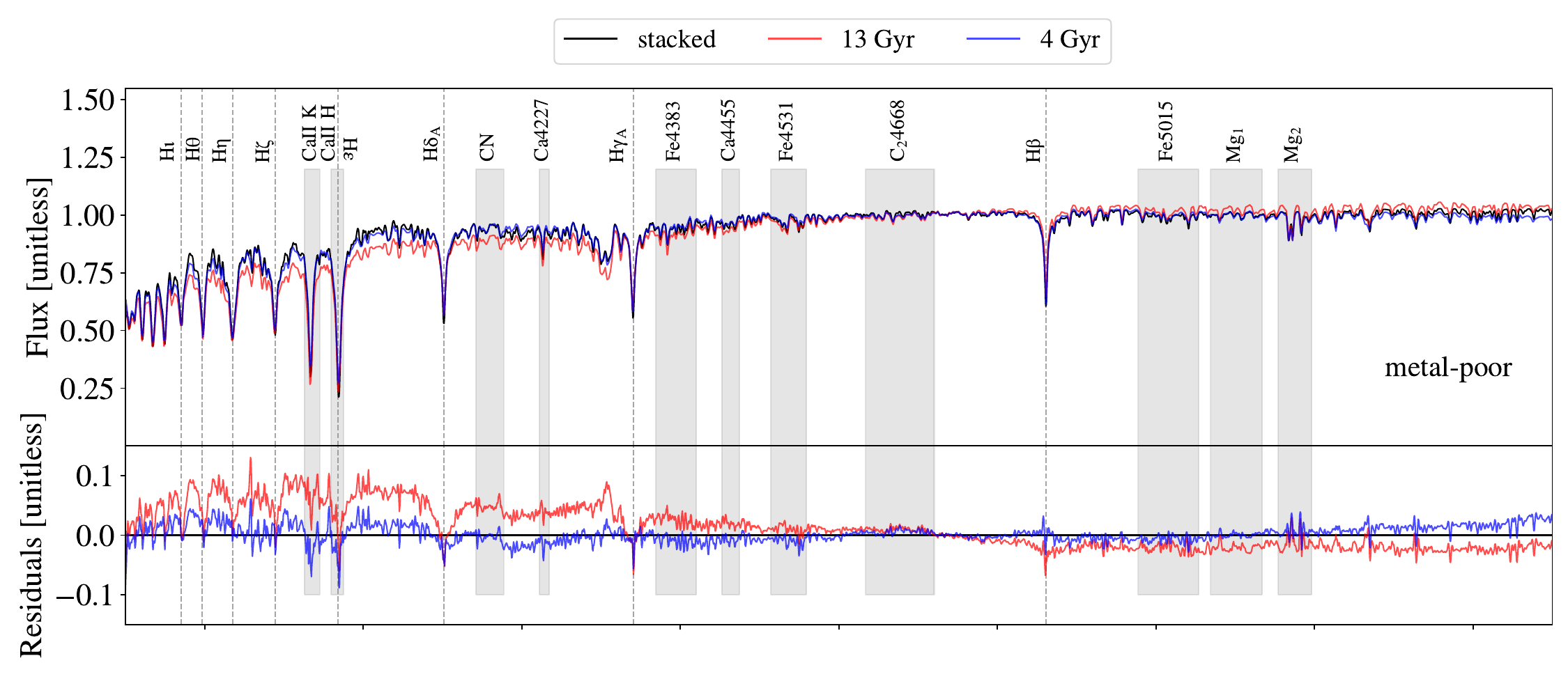} 
         \label{fig:poor}
         \vspace{-0.3cm}
    \end{subfigure}
    \begin{subfigure}[b]{\textwidth}
         \centering
         \includegraphics[width=0.95\textwidth]{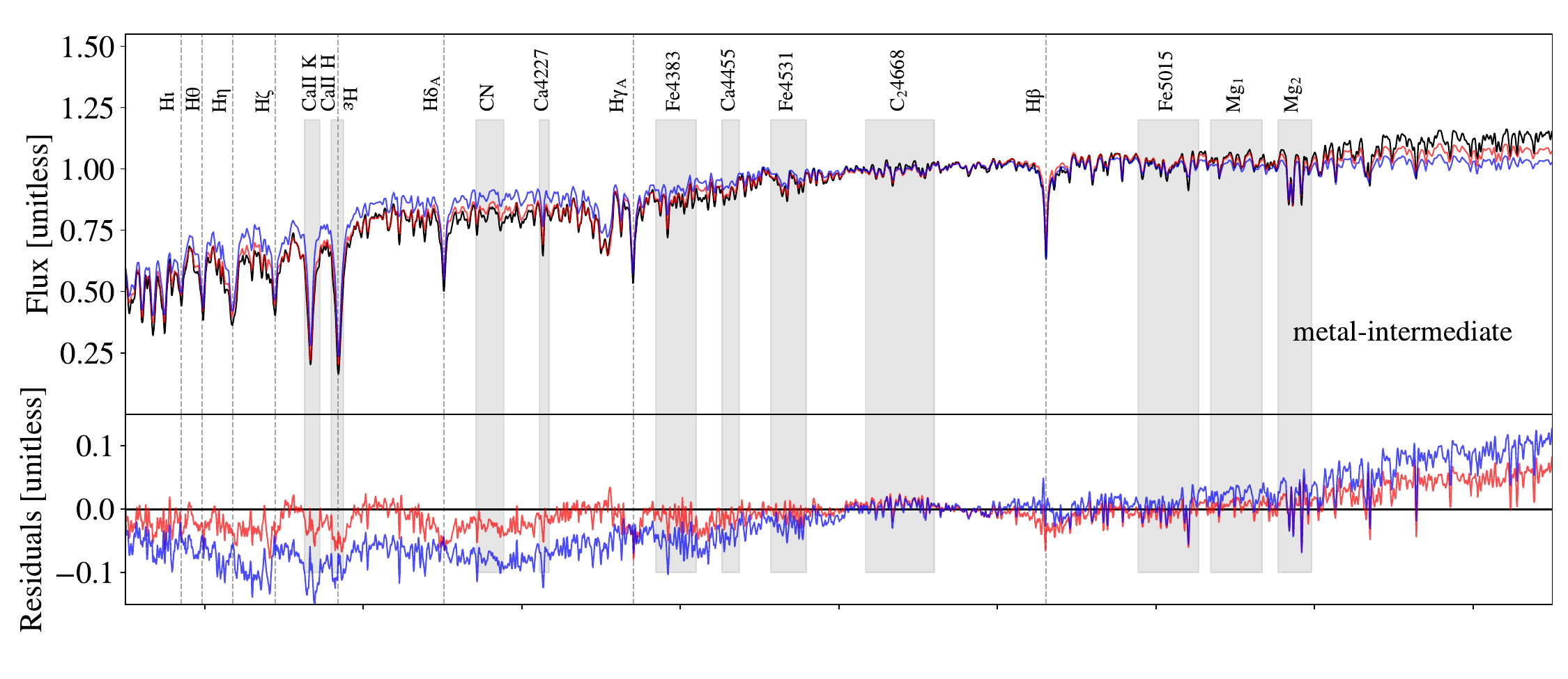}
         \label{fig:intermediate}
         \vspace{-0.5cm}
    \end{subfigure}
    \begin{subfigure}[b]{\textwidth}
         \centering
         \includegraphics[width=0.95\textwidth]{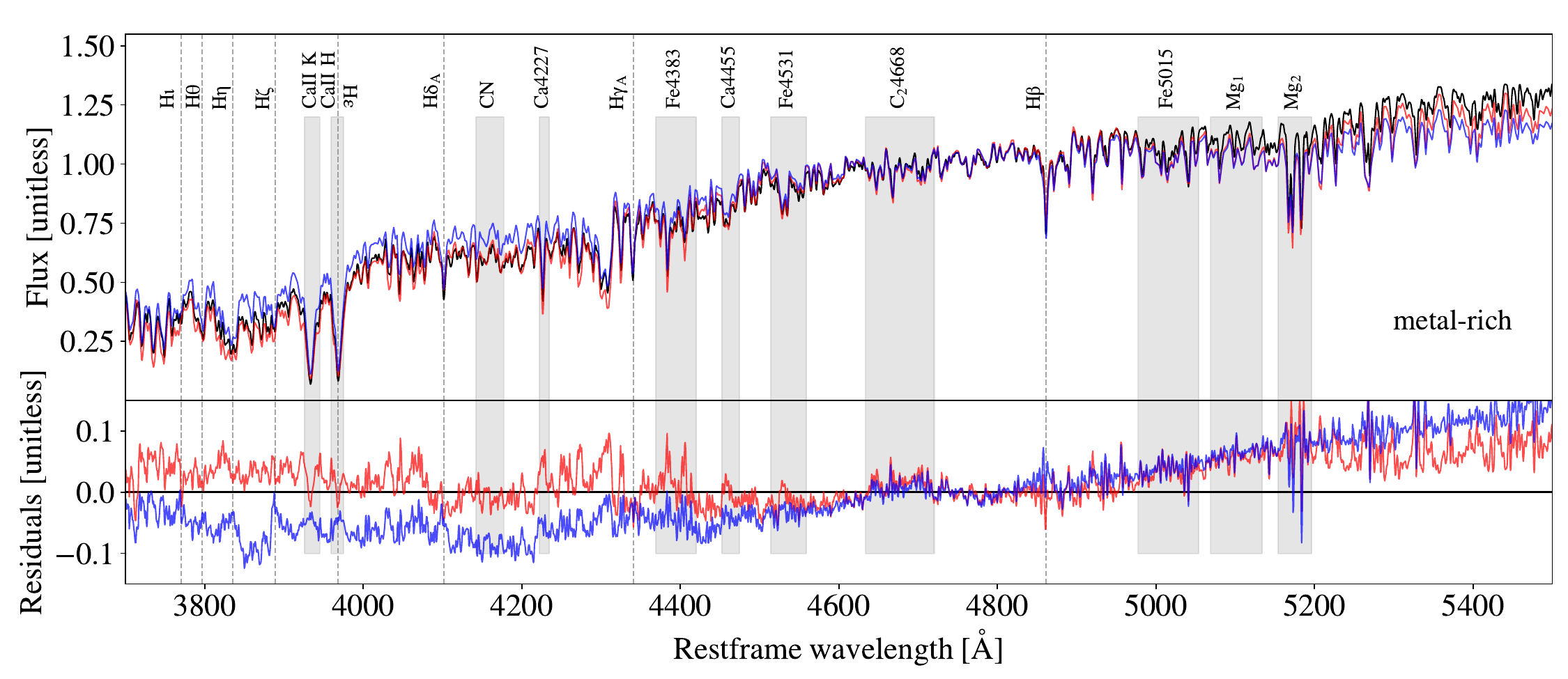}
         \label{fig:rich}
    \end{subfigure}
    \caption{Stacked spectra of the metal-poor (top panel, $\mathrm{[-2.2,-1.4]}$), metal-intermediate (central panel, $\mathrm{[-1.4,-0.6]}$) and metal-rich (bottom panel, $\mathrm{[-0.6,0.2]}$) subsamples, in comparison with a 4 Gyr (in blue) and 13 Gyr old (in red) synthetic spectra. The values of mass, dust absorption and velocity dispersion of the synthetic spectra are set to the median literature values of each subgroup. All the spectra are normalised in the window 4500-5000 {\AA}. At the bottom of each panel the residuals of the stacked spectra with respect to the synthetic ones are shown with corresponding colours. The Balmer series is indicated with dashed lines, other relevant spectral features are highlighted with grey boxes. It can be noticed how the different sub-groups show different continuum and spectral features properties, and how the models are able to reproduce more accurately the high-metallicity ones.}\label{Fig:4stacked}
\end{figure*}

The metal-poor sample counts 15 GCs, among which only the 20\% (3/15) is recognised as older than 10 Gyr. This can be better understood by examining the top panel of Fig. \ref{Fig:4stacked}, where the median stacked spectrum of the metal-poor sample is compared to two synthetic spectra, one 13 Gyr old and the other 4 Gyr old, while all the other parameters (e.g., [Z/H], mass, dust) are fixed to the median literature values characterising this sub-sample. The spectral shape does indeed resemble the one of a young population, with a stronger emission bluer than $\mathrm{\sim4400 \AA}$ with respect to the spectrum of an old population, and stronger Balmer lines. However, in Fig. \ref{fig:CaIIK_Hbeta}, where discrepancies in the reproduction of CaII K and H$\beta$ are shown as a function of literature metallicity, we can see that for most of the metal-poor GCs, the H$\beta$ line is actually overestimated. This means that the young solution, even if preferred by the fit, does not precisely follow the observed features. In terms of metallicity and mass, the agreement with literature values is very good, with average deviations of $\mathrm{\langle \Delta [Z/H] \rangle =- 0.02 \pm 0.35}$ dex and $\mathrm{\langle \Delta \mass \rangle =-0.2 \pm 0.2}$ dex respectively. 

The metal-intermediate is the largest sub-group, with 33 GCs, and shows a higher percentage of GCs older than 10 Gyr compared to the metal-poor sample, equal to 33\% (11/33). Its median stacked spectrum has a better agreement with an old population in terms of the continuum but still fails to reproduce some observed lines of the Balmer series, as we can see in the central panel of Fig. \ref{Fig:4stacked}. This is mostly evident for the H$\beta$ line, falling in the region $\mathrm{4500-5000 \AA}$ where the spectra have been normalised, which is clearly better reproduced by a young population, as in the metal-poor case. This suggests that there is still non-negligible contamination by hot stars, deepening the Balmer series. We can observe this in more detail in Fig. \ref{fig:CaIIK_Hbeta} where to obtain old solutions, the fit has to underestimate the H$\beta$ feature. In contrast, for the young ones, it is either compatible with observations or overestimated. In this metal-intermediate group, we can also observe the importance of reproducing the CaII K line in recovering ages. In fact, while the old solutions all scatter around $\mathrm{\Delta CaII K \sim 0}$, the young ones systematically underestimate this feature. Regarding the metallicity, this sub-group shows a good agreement with literature values, with a $\mathrm{\langle \Delta [Z/H] \rangle =0.10 \pm 0.15}$ dex, and a discrepancy in mass smaller than before $\mathrm{\langle \Delta \mass \rangle = -0.04 \pm 0.30}$ dex.

In the metal-rich sample instead, we are able to obtain ages older than 10 Gyr for 56\% (10/18) of the sample. This fraction increases as we move to solar values, reaching 70\% for $\mathrm{[Z/H]_{lit}\geq -0.4}$. In the bottom panel of Fig. \ref{Fig:4stacked} we can see that, among the three samples, the metal-rich is the most distant from a young population with a much redder continuum. In addition, the metal-rich is the only stacked spectrum in which the old synthetic spectrum accurately reproduces the H$\beta$, while the young one leaves a clear residual. Looking at Fig. \ref{fig:CaIIK_Hbeta}, the metal-rich sample is the only one for which both H$\beta$ and CaII K are well reproduced. In terms of metallicity and mass here we find $\mathrm{\langle \Delta [Z/H] \rangle = -0.08 \pm 0.15}$ dex and $\mathrm{\langle \Delta \mass \rangle =0.09 \pm 0.16}$ dex with respect to literature values.\\
The fact that we can well recover ages in this metallicity range is a remarkable result also in the context of galaxy evolution studies, for which GCs have always been an important test bench, since it shows the reliability of the FSF method in recovering the main physical parameters of the stellar population in a metallicity interval that is typical of high redshift galaxies \citep[see, e.g.,][]{Kriek2019,Lonoce2020,Carnall2022,Borghi2022a}. Moreover, we show that spectral features like H$\beta$ and CaII K prove to be very sensible to possible spurious age determinations and thus can be used as diagnostics to determine the quality of the age estimates.

\subsection{Comparison with previous works}\label{sec:CC22comparison}
As anticipated in Sect. \ref{sec:1Intro}, different works have already been published investigating the potential of analysing GCs' integrated spectra, either relying on different datasets or adopting different fitting codes. Here we focus on the work from \citet{CabreraZiri2022} (CC22, hereafter), where they use the observations from \citet{Schiavon2005} for a common sub-sample of GCs, and then on the results from \citet{Goncalves2020} (G20, hereafter), where they analyse the same data but adopt a non-parametric approach.

\subsubsection{Comparison with Cabrera-Ziri {\&} Conroy (2022)}
\citetalias{CabreraZiri2022} used a similar, but non-parametric, FSF approach with the code ALF \citep{Conroy2012} to estimate the ages and metallicities of 32 Galactic GCs from \citet{Schiavon2005}, fitting normalised spectra in the range $\mathrm{\sim 3300-6500 \AA}$. As in {\it Config. 13.5}, \citetalias{CabreraZiri2022} initially performed the analysis with a standard setting, using a cosmological prior of 14 Gyr. They obtained ages compatible with literature values within 1.5 Gyr for 7 GCs, which constitutes 22\% of their sample. To compare the results, we consider the 31 GCs included both in their sample and our good fits. For those, in {\it Config. 13.5} we obtain ages within 1.5 Gyr from literature values for 13\% of the sample (4/31) while this fraction is more than doubled when we remove the cosmological prior, reaching 32\% (10/31). Compared to \citetalias{CabreraZiri2022}, the performance in recovering old ages with {\it Config. 13.5} is comparable but slightly worse. This is likely due to the additional degrees of freedom in our setting, such as the inclusion of dust and mass parameters, along with the lower age prior. If adopting a multi-component model can help in reproducing all the spectral features in more detail, this approach is also more prone to possible degeneracies, as we underlined in different steps of our analysis. Nevertheless, this same choice allows us to obtain better results when removing the cosmological prior, with an increase in the fraction of old objects of 20\% with respect to our {\it Config. 13.5} and 10\% with respect to the standard setting in \citetalias{CabreraZiri2022}.\\
In \citetalias{CabreraZiri2022}, an additional fit was performed that included a component to account for the hotter fraction of HB stars. This approach allowed them to recover ages older than 10 Gyr for 27 out of 31 GCs, with 24 of these being compatible with literature values within 1.5 Gyr. As already discussed, modelling the HB component is outside the purpose of this work, but represents a promising possibility to be explored in future analyses.  

\subsubsection{Comparison with Goncalves et al. (2020)}
In \citetalias{Goncalves2020} the authors focus on the impact that the wavelength range choice has on the results, adopting the FSF code \texttt{STARLIGHT} \citep{CidFernandes2005}. They fit normalised spectra using MILES SSP models \citep{Vazdekis2015} with ages up to 14 Gyr, [Fe/H] in the range from -2.27 to 0.26 and alpha enhancement either absent or equal to 0.4. Dust reddening is implemented in the code, modelled as in \citet{Cardelli1989}. We compare our results to the ones published in \citet{Goncalves2023}, obtained by fitting the interval 4828 - 5634 {\AA}, a narrow range where the main features detectable are H$\beta$, Mgb triplet, Fe5015, Fe5270 and Fe5335.\\ 
We consider the 58 MW GCs for which we obtain a good fit among the 64 MW GCs published in \citetalias{Goncalves2020}; in this sample, they obtained ages compatible with literature values within 1.5 Gyr for 9 GCs, representing 15\% of the total. In this same sub-group, we have 7 GCs compatible with literature ages in \textit{Config. 13.5}, and 15 in \textit{Config. 15}, corresponding to 12\% and 26\% of the sample, respectively. As in the comparison with \citetalias{CabreraZiri2022}, also in the case of \citetalias{Goncalves2020} our results when the cosmological prior is applied are comparable but slightly worse, and as commented above the reason probably resides in the higher number of parameters involved and the lower age prior. Again, when we remove the cosmological prior, we obtain a major improvement in the fraction of ages compatible with literature, both with respect to our \textit{Config. 13.5} (+14\%) and to \citetalias{Goncalves2020} (+11\%).

It is worth mentioning here that we also tested the impact on the results of fitting the wavelength range adopted in \citetalias{Goncalves2020}, first suggested in \citet{Walcher2009}, and we considered it when computing the systematic uncertainty on ages in Sect. \ref{sec:4syst}. Avoiding all the features bluer than 4828 {\AA}, polluted by the hot HB component, this configuration performs much better for the blue HB, low-metallicity GCs in our sample. In particular, it allows us to recover ages older than 10 Gyr for 63\% of metal-poor GCs and 73\% of metal-intermediate ones. For the metal-rich sample, instead, it yields worse results compared to \textit{Config. 15}, with 47\% of GCs resulting older than 10 Gyr. Nevertheless, the latter sub-group is the one in which the stellar models should be most effective, thanks to the absence of an extended HB component and low alpha-enhancement in the systems, so \textit{Config. 15} was still preferable in terms of the robustness of the results.

\section{Application to cosmology}
\label{sec:5COSMO}
In this section, we analyse what impact our results for GCs' ages, interpreted as lower limits to the age of the Universe $t_U$, can have in the determination of the Hubble constant $H_0$ \citep{Jimenez2019,Valcin2020,Valcin2021,Vagnozzi2022,Cimatti2023}. 

\subsection{Method}\label{sec:5.1Method}
As anticipated in Sect. \ref{sec:1Intro}, and widely described in \citet{Cimatti2023}, $H_0$ is very sensitive to the value of $t_U$. In a generic cosmological model, $H_0$ can be expressed as:
\begin{equation}
    \mathrm{H_0 = \frac{A}{t}\int_0^{z_F}\frac{1}{(1+z)E(z)} dz},
\end{equation}
where t is the age of a given object, $z_F$ its redshift of formation, \textit{E(z)} is defined as \textit{H(z)}/$H_0$ and \textit{A}=977.8 converts the result in units of $\mathrm{\kmsmpc}$. The analytical expression of \textit{E(z)} depends on the cosmological model assumed, in particular, in a flat $\Lambda$CDM model it can be expressed as a function of redshift and matter density parameter $\Omega_M$, so that $H_0$ becomes:
\begin{equation}\label{eq:5H0LCDM}
    \mathrm{H_0 = \frac{A}{t}\int_0^{z_F}\frac{1}{1+z}[\Omega_M(1+z)^3+(1-\Omega_M)]^{-1/2} dz}.
\end{equation}
Considering the limit of $z_F = \infty$ for which $t$ corresponds to the age of the Universe $t_U$, we can easily understand the sensitivity of the method. Fixing $\Omega_M=0.3$ and varying $t_U$ from 12.9 Gyr to 14.1 Gyr the resulting value of $H_0$ spans from 73 $\mathrm{\kmsmpc}$ to 67 $\mathrm{\kmsmpc}$.  

When applying the method to the oldest objects, $H_0$ can be estimated via a Bayesian approach, in which the likelihood is built on the difference between the measured age and the one predicted by the cosmological model ($\mathrm{age_m}$), accounting for the age error ($\mathrm{\sigma_{age}}$): 
\begin{equation}
    \mathrm{\mathcal{L}(age|\textbf{p}) = -0.5 \sum_i \frac{[age_i-age_m(\textbf{p})]^2}{\sigma_{age,i}}},
\end{equation}
where \textbf{p} = ($\mathrm{H_0,z_F,\Omega_M}$) in a flat $\Lambda$CDM cosmology. The posterior distribution then, can be sampled with a Monte Carlo Markov Chain approach like the one implemented in the affine-invariant ensemble sampler \texttt{emcee} \citep{ForemanMackey2013}. In the choice of priors, a flat, uninformative one can be adopted on $H_0$ and $z_F$, while a Gaussian prior is preferable for $\Omega_{M}$ in order to break its intrinsic degeneracy with $H_0$.

As a final note on the method, it is interesting to highlight that nowadays, with facilities like JWST, this approach is no more limited to the study of local objects but can be extended to higher redshift thanks to the first detections of GCs around lensed galaxies. A great case-study is the one of the Sparkler, a galaxy discovered in Webb's First Deep Field \citep{Mowla2022} showing a population of compact objects associated with it. Among the 18 compact objects identified \citep{Mowla2022,Millon2024}, Sparkler shows 5 GC candidates that, if confirmed with spectroscopy, would be the first detection of GCs at z=1.38. The spectroscopic study of these objects, analogous to the one performed in this work, would allow us to measure their age, potentially with a higher precision than the one obtained here for $\sim$13 Gyr old GCs. In younger stellar populations, indeed, the shape of the spectrum and the width of the absorption lines changes much quicker than it does at very old ages, thus reducing the impact of the parameters' degeneracies on the age uncertainty. In the context of our cosmological analysis, extending the method at higher redshift would just require to replace the lower limit of the integral in Eq. \ref{eq:5H0LCDM} with the redshift of the lensed GC. In the case of the Sparkler at redshift 1.38, for example, the age of the Universe ranges from 4.3 Gyr to 4.7 Gyr adopting the reference values for $H_0$ given above, 73 $\mathrm{\kmsmpc}$ and 67 $\mathrm{\kmsmpc}$ respectively.

\subsection{Results}

\begin{figure}
    \centering
    \includegraphics[width=0.45\textwidth]{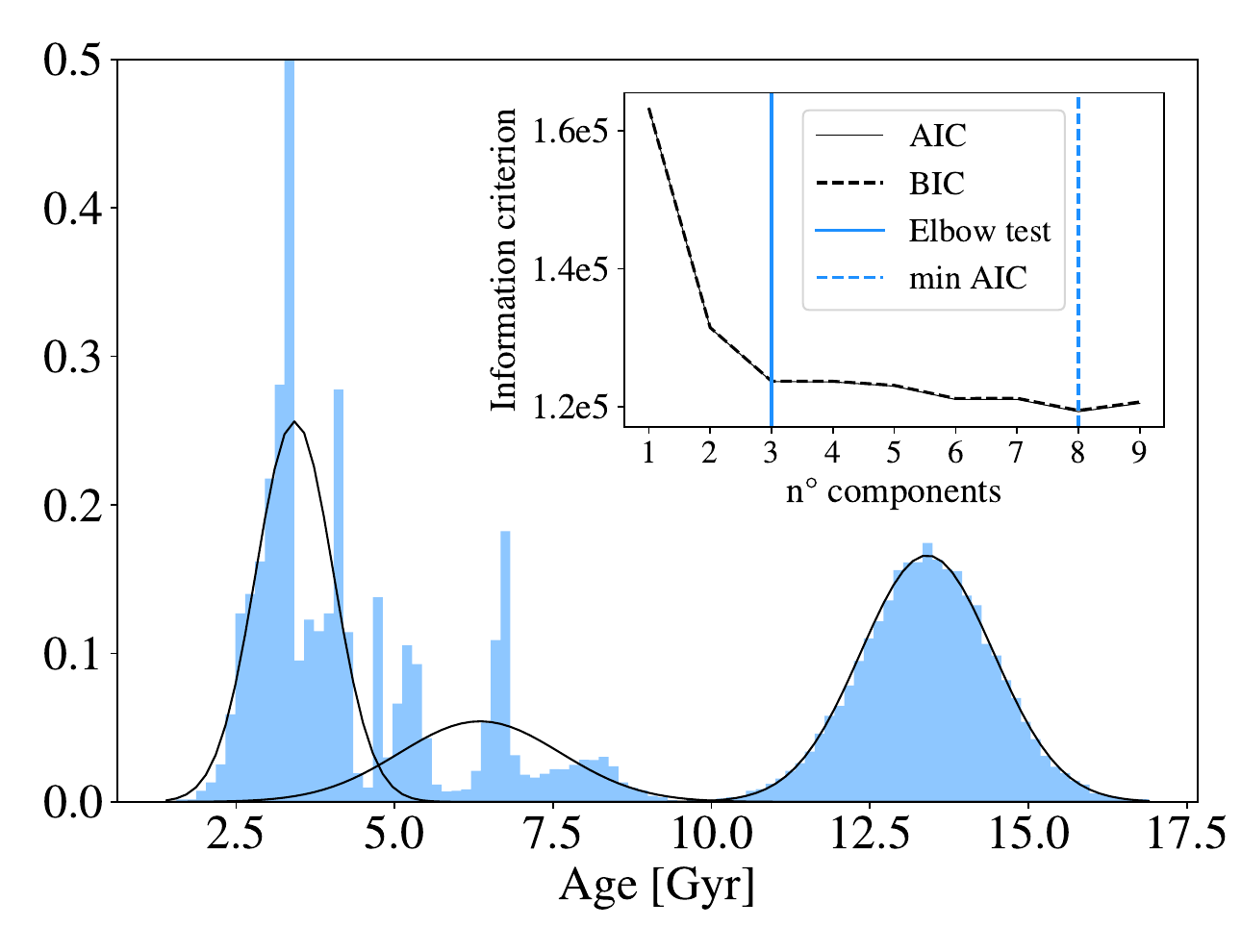}
    \caption{Combination of all the 66 GCs ages, drawn as normal distributions peaked on the best-fit value and 1-$\sigma$ equal to the associated error. In black, the three Gaussian components identified with a GMM are shown. In the inner panel, AIC and BIC curves are shown.}
\end{figure}

As anticipated in Sect. \ref{sec:4results}, we adopted as benchmark \textit{Config. 15}, where the cosmological prior is removed. To identify the tail of the oldest objects, we adopt a Gaussian Mixture Model (GMM) on the whole sample, combining normal distributions peaked on the best-fit ages, with 1-$\sigma$ equal to relative uncertainties. We then let the fit decide the optimal number of subsamples in which to split our data. Both considering the Bayesian Information Criterion (BIC) or the Akaike Information Criterion (AIC), we find that the optimal number of components is three: a first peak identifying the youngest, blue-HB, spurious solutions; a second one for the intermediate ages; a third one comprising all the 24 oldest GCs, peaking at $\mathrm{13.4 \pm 1.1 \:Gyr}$. The latter represents the oldest tail of the GCs' age distribution.

We applied the method described in Sect. \ref{sec:5.1Method} for each of these GCs separately, and then on their average, adopting the following priors: uniform on $\mathrm{H_0 \in [0,150]\: \kmsmpc}$ and on $\mathrm{z_F \in}$ [11,30], Gaussian on $\mathrm{\Omega_{M} = 0.30 \pm 0.02}$. The lower limit on $z_F$ is based on the highest redshift at which galaxies have spectroscopic confirmations \citep{CurtisLake2023}, the higher limit instead relies on the values found in theoretical models for the redshift of formation of the very first stars \citep{Galli2013}. As regards $\Omega_M$, the value chosen here comes from the combination of different low-redshift results \citep{Jimenez2019}, thus independent of the CMB.

Our 24 old GCs span the age range 13.2-13.6 Gyr, with errors around 1 Gyr, thus the resulting $H_0$ range we find covers the interval 69.5-71.7 $\mathrm{\kmsmpc}$, with typical uncertainties of $\sim$5 $\mathrm{\kmsmpc}$. To provide a single $H_0$ measurement, we also ran the MCMC using mean and standard deviation of the old peak found in the GMM fit: $\mathrm{13.4 \pm 1.1 \:Gyr}$. This results in a final value for $H_0$ = $\mathrm{70.4^{+6.7}_{-5.7} \kmsmpc}$ (stat).\\ To account for the systematic contribution to the error budget computed in Sect. \ref{sec:4syst} we summed it in quadrature to the standard deviation of the distribution and ran again the MCMC. The final result, comprising both statistics and systematic effects is: 
\begin{equation*}
    H_0 = \mathrm{70.5^{+7.7}_{-6.3} \kmsmpc \quad (stat + syst).} 
\end{equation*}

\begin{figure}
    \centering
    \includegraphics[width=\linewidth]{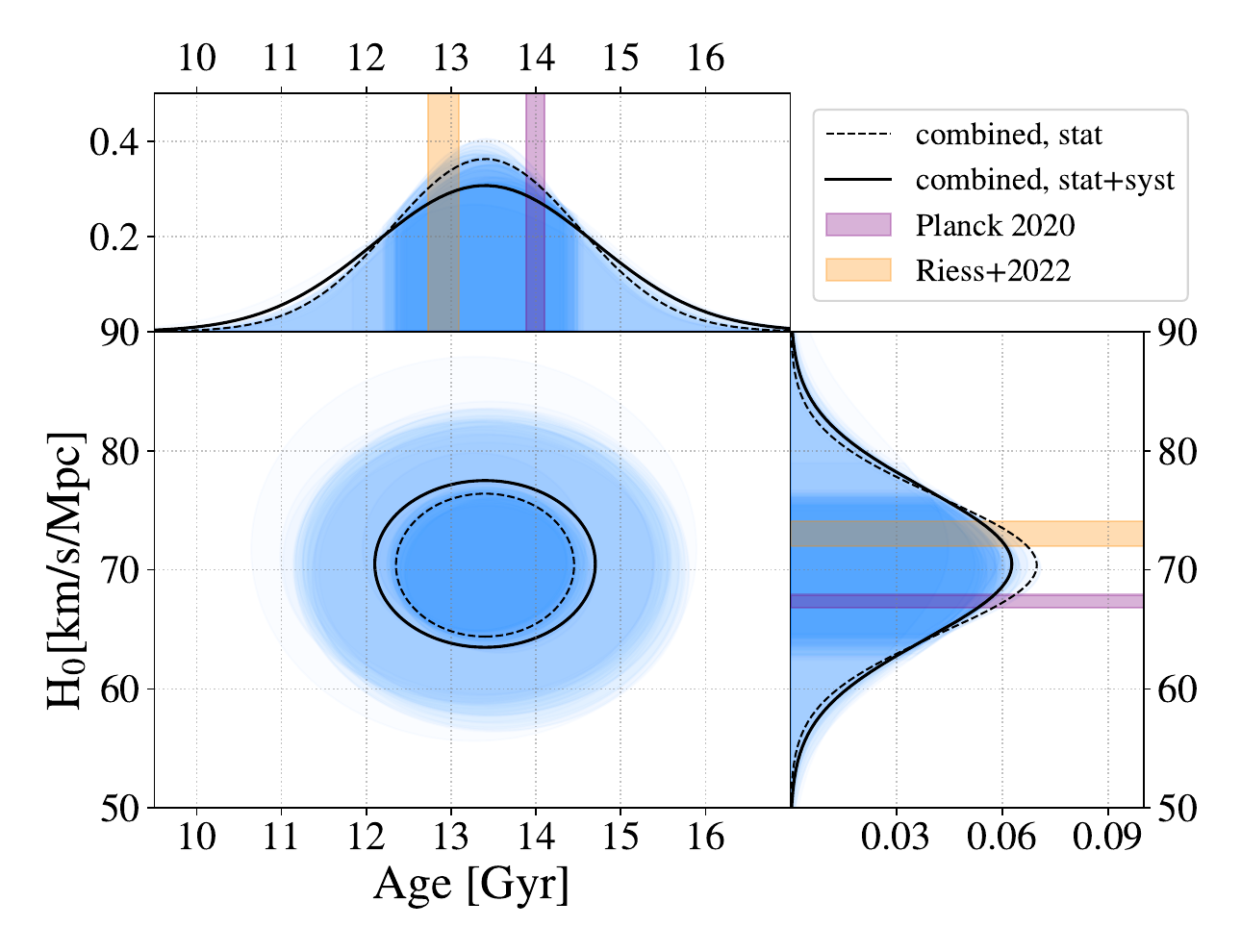}
    \caption{$H_0$ versus age for the sample of 24 old GCs. The ages and corresponding $H_0$ estimates are shown as Gaussians peaked on the best-fit values and 1-$\sigma$ equal to the relative uncertainties. The 1-$\sigma$ limits are also highlighted in a darker blue. The solid black curves correspond to the average GC's age and relative $H_0$ estimate. For comparison, the values from \citet{Riess2022} and \citet{PlanckCollaboration2020} are represented respectively with dashed and dotted lines in the $H_0$ domain, and in the age-domain we report the corresponding ages of the Universe as computed in a flat $\Lambda$CDM with $\mathrm{\Omega_{M}=0.3\: and\: \Omega_{\Lambda}=0.7}$.}
    \label{fig:5H0_age}
\end{figure}

In Fig. \ref{fig:5H0_age} the 24 GCs' ages and the corresponding $H_0$ estimates found after the MCMC run are represented as Gaussian distributions, in the respective domains and combined as ellipses in the $H_0$-age plane. The corresponding Gaussian curves and ellipses relative to the combined age and $H_0$ are shown as black solid lines. For comparison, the values from \citet{Riess2022} and \citet{PlanckCollaboration2020} are represented respectively with dashed and dotted lines.\\
Of course, the results that we obtain here are not able to address the tension but represent a pilot exploration of the use of GCs' ages for cosmological purposes, especially in view of future missions that could potentially discover such objects at higher redshifts.

\section{Conclusions}\label{sec:6CONCLUSIONS}
In this work, we analysed the integrated spectra of a sample of 77 Milky Way GCs from the WAGGS project \citep{Usher2017} and measured their physical properties via FSF with the code \texttt{BAGPIPES} \citep{Carnall2018}. In doing this, we aimed to study how well FSF can recover the GCs' ages and physical parameters, and assess, in particular, how the age estimates are affected by the presence or absence of a cosmological prior. This required a modification on the code, already tested and validated in \citet{Jiao2023} and \citet{Tomasetti2023}, thanks to which a flat non-cosmological prior can be set at 15 Gyr. At the same time, this allowed us to obtain a cosmology-independent lower limit to the age of the Universe, that we used to derive a new constraint on $H_0$, performing a pilot study for future potential applications at higher redshift.

Our results are summarised as follows:
\begin{enumerate}
    \item Measuring age-related spectral features detectable in the spectra, like $D_n4000$ and H${\beta}$, allowed us to build index-age diagrams for these features in different metallicity bins, showing a distribution that well aligns with the trends from theoretical spectral models.
    \item Combining age-related and metallicity-related spectral features we built two different types of index-index diagrams, H$\mathrm{\beta}$-$\mathrm{Mg_b}$ and H$\mathrm{\gamma_F}$-[MgFe]', able to disentangle age and metallicity contributions to the spectral features. Based on how the GCs populated these diagrams, we could have a first estimate of their properties, showing an overall good agreement with the corresponding literature values. Ages, however, appeared underestimated for a fraction of the sample characterised by low metallicity ([Fe/H]<-0.33) and blue HB (HBR>0), that, showing a more prominent H$\beta$, populated the area of ages younger than 8 Gyr. This anticipates how the presence of an unmodelled HB component in the spectra can bias the results toward younger solutions.  
    \item Performing FSF with \texttt{BAGPIPES} we could measure simultaneously the age, metallicity, and mass of the GCs. We tested multiple fit configurations, with different choices of priors, model components, SFHs, and wavelength ranges. Metallicity and mass proved to be very stable against the changes in fit configurations, while age was mostly sensible to the prior limit. In particular, we tested two configurations, one with a cosmological prior set at 13.5 Gyr and another at 15 Gyr (\textit{Config. 15}), thus independent of cosmology. The percentage of GCs for which ages result compatible with the literature values within $\pm$1.5 Gyr increases by 20\% removing the cosmological prior, demonstrating the relevance of this limit in stellar population studies. In \textit{Config. 15} the agreement with literature values is maximum for the sub-group of GCs with [Z/H]>-0.4, the least affected by the presence of blue HBs, reaching 70\%. Metallicity and mass always result well compatible with reference values independently of HBR, [Z/H] or fit setting, with average discrepancies on the whole sample of $\mathrm{\langle \Delta [Z/H] \rangle = -0.02 \pm 0.24}$ dex and $\mathrm{\langle \Delta \mass \rangle = 0.04 \pm 0.28}$, compatible with the typical uncertainties associated with these quantities.
    \item Performing a GMM fit on the whole age distribution as derived in \textit{Config.15}, we identified a tail of 24 old GCs with $\mathrm{\langle age \rangle = 13.4 \pm 1.1\: Gyr}$. In a cosmological framework, this value can be used as a lower limit to the age of the Universe to constrain the Hubble constant. In particular, by fitting the $H_0 - t_U$ relation in a flat $\Lambda$CDM cosmology and assuming $\mathrm{\Omega_m = 0.30 \pm 0.02}$, from low-z measurements, we obtained $\mathrm{H_0 = 70.4_{-5.7}^{+6.7} \:\kmsmpc (stat)}$. Taking into account a systematic contribution to the age error of 0.71 Gyr, based on the age fluctuations in 8 different fit configurations, we obtained: 
\begin{equation*}
    H_0 = \mathrm{70.5^{+7.7}_{-6.3} \:\kmsmpc \quad (stat + syst).} 
\end{equation*}
\end{enumerate}

While we acknowledge that the method adopted in this paper is not intended to compete with other age estimation techniques (e.g., isochrone fitting) for local and resolved objects, it does offer a viable alternative. This study serves as an initial pilot investigation into the feasibility of using only spectroscopic information to determine GC ages, an approach that may be particularly useful for investigating the properties of lensed GCs at higher redshifts, where isochrone fitting is not feasible. Future developments will include expanding the tests to incorporate models with an extended HB. Nonetheless, even without these enhancements, this work provides valuable diagnostics for identifying the most robust and reliable fits.

\begin{acknowledgements}
    We thank Christopher Usher for kindly providing us with WAGGS spectra, Adam Carnall for his help in using \texttt{Bagpipes}, and Frédéric Courbin and Licia Verde for the insightful discussion on the potential of lensed GCs' ages in cosmology. ET acknowledges the support from COST Action CA21136 – “Addressing observational tensions in cosmology with systematics and fundamental physics (CosmoVerse)”, supported by COST (European Cooperation in Science and Technology). Funding for the work of RJ was partially provided by project PID2022-141125NB-I00, and the “Center of Excellence Maria de Maeztu 2020-2023” award to the ICCUB (CEX2019- 000918-M) funded by MCIN/AEI/10.13039/501100011033.  MM acknowledges support from MIUR, PRIN 2022 (grant 2022NY2ZRS 001). MM and AC acknowledge support from the grant ASI n. 2024-10-HH.0 “Attività scientifiche per la missione Euclid – fase E”.
    
\end{acknowledgements}

\bibliographystyle{aa}
\bibliography{references}

\begin{thebibliography}{93}
\expandafter\ifx\csname natexlab\endcsname\relax\def\natexlab#1{#1}\fi

\bibitem[{{Abdalla} {et~al.}(2022){Abdalla}, {Abell{\'a}n}, {Aboubrahim}, {Agnello}, {Akarsu}, {Akrami}, {Alestas}, {Aloni}, {Amendola}, {Anchordoqui}, {Anderson}, {Arendse}, {Asgari}, {Ballardini}, {Barger}, {Basilakos}, {Batista}, {Battistelli}, {Battye}, {Benetti}, {Benisty}, {Berlin}, {de Bernardis}, {Berti}, {Bidenko}, {Birrer}, {Blakeslee}, {Boddy}, {Bom}, {Bonilla}, {Borghi}, {Bouchet}, {Braglia}, {Buchert}, {Buckley-Geer}, {Calabrese}, {Caldwell}, {Camarena}, {Capozziello}, {Casertano}, {Chen}, {Chluba}, {Chen}, {Chen}, {Chudaykin}, {Cicoli}, {Copi}, {Courbin}, {Cyr-Racine}, {Czerny}, {Dainotti}, {D'Amico}, {Davis}, {de Cruz P{\'e}rez}, {de Haro}, {Delabrouille}, {Denton}, {Dhawan}, {Dienes}, {Di Valentino}, {Du}, {Eckert}, {Escamilla-Rivera}, {Fert{\'e}}, {Finelli}, {Fosalba}, {Freedman}, {Frusciante}, {Gazta{\~n}aga}, {Giar{\`e}}, {Giusarma}, {G{\'o}mez-Valent}, {Handley}, {Harrison}, {Hart}, {Hazra}, {Heavens}, {Heinesen}, {Hildebrandt}, {Hill}, {Hogg}, {Holz}, {Hooper}, {Hosseininejad}, {Huterer},
  {Ishak}, {Ivanov}, {Jaffe}, {Jang}, {Jedamzik}, {Jimenez}, {Joseph}, {Joudaki}, {Kamionkowski}, {Karwal}, {Kazantzidis}, {Keeley}, {Klasen}, {Komatsu}, {Koopmans}, {Kumar}, {Lamagna}, {Lazkoz}, {Lee}, {Lesgourgues}, {Levi Said}, {Lewis}, {L'Huillier}, {Lucca}, {Maartens}, {Macri}, {Marfatia}, {Marra}, {Martins}, {Masi}, {Matarrese}, {Mazumdar}, {Melchiorri}, {Mena}, {Mersini-Houghton}, {Mertens}, {Milakovi{\'c}}, {Minami}, {Miranda}, {Moreno-Pulido}, {Moresco}, {Mota}, {Mottola}, {Mozzon}, {Muir}, {Mukherjee}, {Mukherjee}, {Naselsky}, {Nath}, {Nesseris}, {Niedermann}, {Notari}, {Nunes}, {{\'O} Colg{\'a}in}, {Owens}, {{\"O}z{\"u}lker}, {Pace}, {Paliathanasis}, {Palmese}, {Pan}, {Paoletti}, {Perez Bergliaffa}, {Perivolaropoulos}, {Pesce}, {Pettorino}, {Philcox}, {Pogosian}, {Poulin}, {Poulot}, {Raveri}, {Reid}, {Renzi}, {Riess}, {Sabla}, {Salucci}, {Salzano}, {Saridakis}, {Sathyaprakash}, {Schmaltz}, {Sch{\"o}neberg}, {Scolnic}, {Sen}, {Sehgal}, {Shafieloo}, {Sheikh-Jabbari}, {Silk}, {Silvestri}, {Skara},
  {Sloth}, {Soares-Santos}, {Sol{\`a} Peracaula}, {Songsheng}, {Soriano}, {Staicova}, {Starkman}, {Szapudi}, {Teixeira}, {Thomas}, {Treu}, {Trott}, {van de Bruck}, {Vazquez}, {Verde}, {Visinelli}, {Wang}, {Wang}, {Wang}, {Watkins}, {Watson}, {Webb}, {Weiner}, {Weltman}, {Witte}, {Wojtak}, {Yadav}, {Yang}, {Zhao}, \& {Zumalac{\'a}rregui}}]{Abdalla2022}
{Abdalla}, E., {Abell{\'a}n}, G.~F., {Aboubrahim}, A., {et~al.} 2022, Journal of High Energy Astrophysics, 34, 49

\bibitem[{Annibali {et~al.}(2018)Annibali, Morandi, Watkins, Tosi, Aloisi, Buzzoni, Cusano, Fumana, Marchetti, Mignoli, Mucciarelli, Romano, \& van~der Marel}]{Annibali2018}
Annibali, F., Morandi, E., Watkins, L.~L., {et~al.} 2018, \mnras, 476, 1942

\bibitem[{Bastian \& Lardo(2018)}]{Bastian2018}
Bastian, N. \& Lardo, C. 2018, \araa, 56, 83

\bibitem[{Baumgardt {et~al.}(2023)Baumgardt, H{\'e}nault-Brunet, Dickson, \& Sollima}]{Baumgardt2023}
Baumgardt, H., H{\'e}nault-Brunet, V., Dickson, N., \& Sollima, A. 2023, \mnras, 521, 3991

\bibitem[{Bennett {et~al.}(2003)Bennett, Halpern, Hinshaw, Jarosik, Kogut, Limon, Meyer, Page, Spergel, Tucker, Wollack, Wright, Barnes, Greason, Hill, Komatsu, Nolta, Odegard, Peiris, Verde, \& Weiland}]{Bennett2003}
Bennett, C.~L., Halpern, M., Hinshaw, G., {et~al.} 2003, \apjs, 148, 1

\bibitem[{Borghi {et~al.}(2022a)Borghi, Moresco, Cimatti, Huchet, Quai, \& Pozzetti}]{Borghi2022a}
Borghi, N., Moresco, M., Cimatti, A., {et~al.} 2022a, \apj, 927, 164

\bibitem[{Brown {et~al.}(2018)Brown, Casertano, Strader, Riess, VandenBerg, Soderblom, Kalirai, \& Salinas}]{Brown2018}
Brown, T.~M., Casertano, S., Strader, J., {et~al.} 2018, \apjl, 856, L6

\bibitem[{Bruzual \& Charlot(2003)}]{Bruzual2003}
Bruzual, G. \& Charlot, S. 2003, \mnras, 344, 1000

\bibitem[{Buchner(2016)}]{Buchner2016}
Buchner, J. 2016, Statistics and Computing, 26, 383

\bibitem[{Burstein {et~al.}(1984)Burstein, Faber, Gaskell, \& Krumm}]{Burstein1984}
Burstein, D., Faber, S.~M., Gaskell, C.~M., \& Krumm, N. 1984, \apj, 287, 586

\bibitem[{Cabrera-Ziri \& Conroy(2022)}]{CabreraZiri2022}
Cabrera-Ziri, I. \& Conroy, C. 2022, \mnras, 511, 341

\bibitem[{Calzetti {et~al.}(2000)Calzetti, Armus, Bohlin, Kinney, Koornneef, \& Storchi-Bergmann}]{Calzetti2000}
Calzetti, D., Armus, L., Bohlin, R.~C., {et~al.} 2000, \apj, 533, 682

\bibitem[{Cardelli {et~al.}(1989)Cardelli, Clayton, \& Mathis}]{Cardelli1989}
Cardelli, J.~A., Clayton, G.~C., \& Mathis, J.~S. 1989, \apj, 345, 245

\bibitem[{Carnall {et~al.}(2019)Carnall, McLure, Dunlop, Cullen, McLeod, Wild, Johnson, Appleby, Dav{\'e}, Amorin, Bolzonella, Castellano, Cimatti, Cucciati, Gargiulo, Garilli, Marchi, Pentericci, Pozzetti, Schreiber, Talia, \& Zamorani}]{Carnall2019}
Carnall, A.~C., McLure, R.~J., Dunlop, J.~S., {et~al.} 2019, \mnras, 490, 417

\bibitem[{Carnall {et~al.}(2018)Carnall, McLure, Dunlop, \& Dav{\'e}}]{Carnall2018}
Carnall, A.~C., McLure, R.~J., Dunlop, J.~S., \& Dav{\'e}, R. 2018, \mnras, 480, 4379

\bibitem[{Carnall {et~al.}(2022)Carnall, McLure, Dunlop, Hamadouche, Cullen, McLeod, Begley, Amorin, Bolzonella, Castellano, Cimatti, Fontanot, Gargiulo, Garilli, Mannucci, Pentericci, Talia, Zamorani, Calabro, Cresci, \& Hathi}]{Carnall2022}
Carnall, A.~C., McLure, R.~J., Dunlop, J.~S., {et~al.} 2022, \apj, 929, 131

\bibitem[{Cassisi \& Salaris(2013)}]{Cassisi2013}
Cassisi, S. \& Salaris, M. 2013, Old Stellar Populations: How to Study the Fossil Record of Galaxy Formation

\bibitem[{Cezario {et~al.}(2013)Cezario, Coelho, Alves-Brito, Forbes, \& Brodie}]{Cezario2013}
Cezario, E., Coelho, P.~R.~T., Alves-Brito, A., Forbes, D.~A., \& Brodie, J.~P. 2013, \aap, 549, A60

\bibitem[{Chevallard \& Charlot(2016)}]{Chevallard2016}
Chevallard, J. \& Charlot, S. 2016, \mnras, 462, 1415

\bibitem[{Chilingarian {et~al.}(2007)Chilingarian, Prugniel, Sil'chenko, \& Koleva}]{Chilingarian2007}
Chilingarian, I., Prugniel, P., Sil'chenko, O., \& Koleva, M. 2007, in IAU Symposium, Vol. 241, Stellar Populations as Building Blocks of Galaxies, ed. A.~{Vazdekis} \& R.~{Peletier}, 175--176

\bibitem[{Cid~Fernandes {et~al.}(2005)Cid~Fernandes, Mateus, Sodr{\'e}, Stasi{\'n}ska, \& Gomes}]{CidFernandes2005}
Cid~Fernandes, R., Mateus, A., Sodr{\'e}, L., Stasi{\'n}ska, G., \& Gomes, J.~M. 2005, \mnras, 358, 363

\bibitem[{Cimatti \& Moresco(2023)}]{Cimatti2023}
Cimatti, A. \& Moresco, M. 2023, \apj, 953, 149

\bibitem[{Colucci {et~al.}(2017)Colucci, Bernstein, \& McWilliam}]{Colucci2017}
Colucci, J.~E., Bernstein, R.~A., \& McWilliam, A. 2017, \apj, 834, 105

\bibitem[{Conroy \& van Dokkum(2012)}]{Conroy2012}
Conroy, C. \& van Dokkum, P. 2012, \apj, 747, 69

\bibitem[{Curtis-Lake {et~al.}(2023)Curtis-Lake, Carniani, Cameron, Charlot, Jakobsen, Maiolino, Bunker, Witstok, Smit, Chevallard, Willott, Ferruit, Arribas, Bonaventura, Curti, D'Eugenio, Franx, Giardino, Looser, L{\"u}tzgendorf, Maseda, Rawle, Rix, Rodr{\'\i}guez~del Pino, {\"U}bler, Sirianni, Dressler, Egami, Eisenstein, Endsley, Hainline, Hausen, Johnson, Rieke, Robertson, Shivaei, Stark, Tacchella, Williams, Willmer, Bhatawdekar, Bowler, Boyett, Chen, de~Graaff, Helton, Hviding, Jones, Kumari, Lyu, Nelson, Perna, Sandles, Saxena, Suess, Sun, Topping, Wallace, \& Whitler}]{CurtisLake2023}
Curtis-Lake, E., Carniani, S., Cameron, A., {et~al.} 2023, Nature Astronomy, 7, 622

\bibitem[{Eisenstein {et~al.}(2005)Eisenstein, Zehavi, Hogg, Scoccimarro, Blanton, Nichol, Scranton, Seo, Tegmark, Zheng, Anderson, Annis, Bahcall, Brinkmann, Burles, Castander, Connolly, Csabai, Doi, Fukugita, Frieman, Glazebrook, Gunn, Hendry, Hennessy, Ivezi{\'c}, Kent, Knapp, Lin, Loh, Lupton, Margon, McKay, Meiksin, Munn, Pope, Richmond, Schlegel, Schneider, Shimasaku, Stoughton, Strauss, SubbaRao, Szalay, Szapudi, Tucker, Yanny, \& York}]{Eisenstein2005}
Eisenstein, D.~J., Zehavi, I., Hogg, D.~W., {et~al.} 2005, \apj, 633, 560

\bibitem[{Faber {et~al.}(1985)Faber, Friel, Burstein, \& Gaskell}]{Faber1985}
Faber, S.~M., Friel, E.~D., Burstein, D., \& Gaskell, C.~M. 1985, \apjs, 57, 711

\bibitem[{Foreman-Mackey {et~al.}(2013)Foreman-Mackey, Hogg, Lang, \& Goodman}]{ForemanMackey2013}
Foreman-Mackey, D., Hogg, D.~W., Lang, D., \& Goodman, J. 2013, \pasp, 125, 306

\bibitem[{Galli \& Palla(2013)}]{Galli2013}
Galli, D. \& Palla, F. 2013, \araa, 51, 163

\bibitem[{Gon{\c{c}}alves {et~al.}(2020)Gon{\c{c}}alves, Coelho, Schiavon, \& Usher}]{Goncalves2020}
Gon{\c{c}}alves, G., Coelho, P., Schiavon, R., \& Usher, C. 2020, \mnras, 499, 2327

\bibitem[{Goncalves {et~al.}(2023)Goncalves, Coelho, Schiavon, \& Usher}]{Goncalves2023}
Goncalves, G., Coelho, P., Schiavon, R., \& Usher, C. 2023, VizieR Online Data Catalog, J/MNRAS/499/2327

\bibitem[{Gratton {et~al.}(2019)Gratton, Bragaglia, Carretta, D'Orazi, Lucatello, \& Sollima}]{Gratton2019}
Gratton, R., Bragaglia, A., Carretta, E., {et~al.} 2019, \aapr, 27, 8

\bibitem[{Harris(1996)}]{Harris1996}
Harris, W.~E. 1996, \aj, 112, 1487

\bibitem[{Harris(2010)}]{Harris2010}
Harris, W.~E. 2010, arXiv e-prints, arXiv:1012.3224

\bibitem[{H{\'e}nault-Brunet {et~al.}(2019)H{\'e}nault-Brunet, Gieles, Sollima, Watkins, Zocchi, Claydon, Pancino, \& Baumgardt}]{HenaultBrunet2019}
H{\'e}nault-Brunet, V., Gieles, M., Sollima, A., {et~al.} 2019, \mnras, 483, 1400

\bibitem[{Jiao {et~al.}(2023)Jiao, Borghi, Moresco, \& Zhang}]{Jiao2023}
Jiao, K., Borghi, N., Moresco, M., \& Zhang, T.-J. 2023, \apjs, 265, 48

\bibitem[{Jimenez {et~al.}(2019)Jimenez, Cimatti, Verde, Moresco, \& Wandelt}]{Jimenez2019}
Jimenez, R., Cimatti, A., Verde, L., Moresco, M., \& Wandelt, B. 2019, \jcap, 2019, 043

\bibitem[{Jimenez \& Loeb(2002)}]{Jimenez2002}
Jimenez, R. \& Loeb, A. 2002, \apj, 573, 37

\bibitem[{Jimenez {et~al.}(2004)Jimenez, MacDonald, Dunlop, Padoan, \& Peacock}]{Jimenez2004}
Jimenez, R., MacDonald, J., Dunlop, J.~S., Padoan, P., \& Peacock, J.~A. 2004, \mnras, 349, 240

\bibitem[{Koleva {et~al.}(2009)Koleva, Prugniel, Bouchard, \& Wu}]{Koleva2009}
Koleva, M., Prugniel, P., Bouchard, A., \& Wu, Y. 2009, \aap, 501, 1269

\bibitem[{Koleva {et~al.}(2008)Koleva, Prugniel, Ocvirk, Le~Borgne, \& Soubiran}]{Koleva2008}
Koleva, M., Prugniel, P., Ocvirk, P., Le~Borgne, D., \& Soubiran, C. 2008, \mnras, 385, 1998

\bibitem[{Kriek {et~al.}(2019)Kriek, Price, Conroy, Suess, Mowla, Pasha, Bezanson, van Dokkum, \& Barro}]{Kriek2019}
Kriek, M., Price, S.~H., Conroy, C., {et~al.} 2019, \apjl, 880, L31

\bibitem[{Kroupa(2001)}]{Kroupa2001}
Kroupa, P. 2001, \mnras, 322, 231

\bibitem[{Lee {et~al.}(2000)Lee, Yoon, \& Lee}]{Lee2000}
Lee, H.-c., Yoon, S.-J., \& Lee, Y.-W. 2000, \aj, 120, 998

\bibitem[{Lee(1989)}]{Lee1989}
Lee, Y.-W. 1989, PhD thesis, Yale University, Connecticut

\bibitem[{Lee {et~al.}(1994)Lee, Demarque, \& Zinn}]{Lee1994}
Lee, Y.-W., Demarque, P., \& Zinn, R. 1994, \apj, 423, 248

\bibitem[{Lonoce {et~al.}(2020)Lonoce, Maraston, Thomas, Longhetti, Parikh, Guarnieri, \& Comparat}]{Lonoce2020}
Lonoce, I., Maraston, C., Thomas, D., {et~al.} 2020, \mnras, 492, 326

\bibitem[{Maraston \& Str{\"o}mb{\"a}ck(2011)}]{Maraston2011}
Maraston, C. \& Str{\"o}mb{\"a}ck, G. 2011, \mnras, 418, 2785

\bibitem[{Mendel {et~al.}(2007)Mendel, Proctor, \& Forbes}]{Mendel2007}
Mendel, J.~T., Proctor, R.~N., \& Forbes, D.~A. 2007, \mnras, 379, 1618

\bibitem[{Millon {et~al.}(2024)Millon, Michalewicz, Dux, Courbin, \& Marshall}]{Millon2024}
Millon, M., Michalewicz, K., Dux, F., Courbin, F., \& Marshall, P.~J. 2024, \aj, 168, 55

\bibitem[{Milone \& Marino(2022)}]{Milone2022}
Milone, A.~P. \& Marino, A.~F. 2022, Universe, 8, 359

\bibitem[{Moresco(2015)}]{Moresco2015}
Moresco, M. 2015, \mnras, 450, L16

\bibitem[{Moresco {et~al.}(2022)Moresco, Amati, Amendola, Birrer, Blakeslee, Cantiello, Cimatti, Darling, Della~Valle, Fishbach, Grillo, Hamaus, Holz, Izzo, Jimenez, Lusso, Meneghetti, Piedipalumbo, Pisani, Pourtsidou, Pozzetti, Quartin, Risaliti, Rosati, \& Verde}]{Moresco2022}
Moresco, M., Amati, L., Amendola, L., {et~al.} 2022, Living Reviews in Relativity, 25, 6

\bibitem[{Moresco {et~al.}(2012)Moresco, Cimatti, Jimenez, Pozzetti, Zamorani, Bolzonella, Dunlop, Lamareille, Mignoli, Pearce, Rosati, Stern, Verde, Zucca, Carollo, Contini, Kneib, Le~F{\`e}vre, Lilly, Mainieri, Renzini, Scodeggio, Balestra, Gobat, McLure, Bardelli, Bongiorno, Caputi, Cucciati, de~la Torre, de~Ravel, Franzetti, Garilli, Iovino, Kampczyk, Knobel, Kova{\v{c}}, Le~Borgne, Le~Brun, Maier, Pell{\'o}, Peng, Perez-Montero, Presotto, Silverman, Tanaka, Tasca, Tresse, Vergani, Almaini, Barnes, Bordoloi, Bradshaw, Cappi, Chuter, Cirasuolo, Coppa, Diener, Foucaud, Hartley, Kamionkowski, Koekemoer, L{\'o}pez-Sanjuan, McCracken, Nair, Oesch, Stanford, \& Welikala}]{Moresco2012}
Moresco, M., Cimatti, A., Jimenez, R., {et~al.} 2012, \jcap, 2012, 006

\bibitem[{Moresco {et~al.}(2016)Moresco, Pozzetti, Cimatti, Jimenez, Maraston, Verde, Thomas, Citro, Tojeiro, \& Wilkinson}]{Moresco2016}
Moresco, M., Pozzetti, L., Cimatti, A., {et~al.} 2016, \jcap, 2016, 014

\bibitem[{Mowla {et~al.}(2022)Mowla, Iyer, Desprez, Estrada-Carpenter, Martis, Noirot, Sarrouh, Strait, Asada, Abraham, Brammer, Sawicki, Willott, Bradac, Doyon, Muzzin, Pacifici, Ravindranath, \& Zabl}]{Mowla2022}
Mowla, L., Iyer, K.~G., Desprez, G., {et~al.} 2022, \apjl, 937, L35

\bibitem[{Ocvirk {et~al.}(2006)Ocvirk, Pichon, Lan{\c{c}}on, \& Thi{\'e}baut}]{Ocvirk2006}
Ocvirk, P., Pichon, C., Lan{\c{c}}on, A., \& Thi{\'e}baut, E. 2006, \mnras, 365, 74

\bibitem[{Oliveira {et~al.}(2020)Oliveira, Souza, Kerber, Barbuy, Ortolani, Piotto, Nardiello, P{\'e}rez-Villegas, Maia, Bica, Cassisi, D'Antona, Lagioia, Libralato, Milone, Anderson, Aparicio, Bedin, Brown, King, Marino, Pietrinferni, Renzini, Sarajedini, van~der Marel, \& Vesperini}]{Oliveira2020}
Oliveira, R.~A.~P., Souza, S.~O., Kerber, L.~O., {et~al.} 2020, \apj, 891, 37

\bibitem[{O'Malley {et~al.}(2017)O'Malley, Gilligan, \& Chaboyer}]{OMalley2017}
O'Malley, E.~M., Gilligan, C., \& Chaboyer, B. 2017, \apj, 838, 162

\bibitem[{Onodera {et~al.}(2015)Onodera, Carollo, Renzini, Cappellari, Mancini, Arimoto, Daddi, Gobat, Strazzullo, Tacchella, \& Yamada}]{Onodera2015}
Onodera, M., Carollo, C.~M., Renzini, A., {et~al.} 2015, \apj, 808, 161

\bibitem[{Pagel \& Tautvaisiene(1995)}]{Pagel1995}
Pagel, B.~E.~J. \& Tautvaisiene, G. 1995, \mnras, 276, 505

\bibitem[{Pancino {et~al.}(2017)Pancino, Romano, Tang, Tautvai{\v{s}}ien{\.{e}}, Casey, Gruyters, Geisler, San~Roman, Randich, Alfaro, Bragaglia, Flaccomio, Korn, Recio-Blanco, Smiljanic, Carraro, Bayo, Costado, Damiani, Jofr{\'e}, Lardo, de~Laverny, Monaco, Morbidelli, Sbordone, Sousa, \& Villanova}]{Pancino2017}
Pancino, E., Romano, D., Tang, B., {et~al.} 2017, \aap, 601, A112

\bibitem[{Percival \& Salaris(2011)}]{Percival2011}
Percival, S.~M. \& Salaris, M. 2011, \mnras, 412, 2445

\bibitem[{Percival {et~al.}(2001)Percival, Baugh, Bland-Hawthorn, Bridges, Cannon, Cole, Colless, Collins, Couch, Dalton, De~Propris, Driver, Efstathiou, Ellis, Frenk, Glazebrook, Jackson, Lahav, Lewis, Lumsden, Maddox, Moody, Norberg, Peacock, Peterson, Sutherland, \& Taylor}]{Percival2001}
Percival, W.~J., Baugh, C.~M., Bland-Hawthorn, J., {et~al.} 2001, \mnras, 327, 1297

\bibitem[{Perlmutter {et~al.}(1999)Perlmutter, Aldering, Goldhaber, Knop, Nugent, Castro, Deustua, Fabbro, Goobar, Groom, Hook, Kim, Kim, Lee, Nunes, Pain, Pennypacker, Quimby, Lidman, Ellis, Irwin, McMahon, Ruiz-Lapuente, Walton, Schaefer, Boyle, Filippenko, Matheson, Fruchter, Panagia, Newberg, Couch, \& Project}]{Perlmutter1999}
Perlmutter, S., Aldering, G., Goldhaber, G., {et~al.} 1999, \apj, 517, 565

\bibitem[{{Planck Collaboration} {et~al.}(2020){Planck Collaboration}, Aghanim, Akrami, Ashdown, Aumont, Baccigalupi, Ballardini, Banday, Barreiro, Bartolo, Basak, Battye, Benabed, Bernard, Bersanelli, Bielewicz, Bock, Bond, Borrill, Bouchet, Boulanger, Bucher, Burigana, Butler, Calabrese, Cardoso, Carron, Challinor, Chiang, Chluba, Colombo, Combet, Contreras, Crill, Cuttaia, de~Bernardis, de~Zotti, Delabrouille, Delouis, Di~Valentino, Diego, Dor{\'e}, Douspis, Ducout, Dupac, Dusini, Efstathiou, Elsner, En{\ss}lin, Eriksen, Fantaye, Farhang, Fergusson, Fernandez-Cobos, Finelli, Forastieri, Frailis, Fraisse, Franceschi, Frolov, Galeotta, Galli, Ganga, G{\'e}nova-Santos, Gerbino, Ghosh, Gonz{\'a}lez-Nuevo, G{\'o}rski, Gratton, Gruppuso, Gudmundsson, Hamann, Handley, Hansen, Herranz, Hildebrandt, Hivon, Huang, Jaffe, Jones, Karakci, Keih{\"a}nen, Keskitalo, Kiiveri, Kim, Kisner, Knox, Krachmalnicoff, Kunz, Kurki-Suonio, Lagache, Lamarre, Lasenby, Lattanzi, Lawrence, Le~Jeune, Lemos, Lesgourgues, Levrier, Lewis,
  Liguori, Lilje, Lilley, Lindholm, L{\'o}pez-Caniego, Lubin, Ma, Mac{\'\i}as-P{\'e}rez, Maggio, Maino, Mandolesi, Mangilli, Marcos-Caballero, Maris, Martin, Martinelli, Mart{\'\i}nez-Gonz{\'a}lez, Matarrese, Mauri, McEwen, Meinhold, Melchiorri, Mennella, Migliaccio, Millea, Mitra, Miville-Desch{\^e}nes, Molinari, Montier, Morgante, Moss, Natoli, N{\o}rgaard-Nielsen, Pagano, Paoletti, Partridge, Patanchon, Peiris, Perrotta, Pettorino, Piacentini, Polastri, Polenta, Puget, Rachen, Reinecke, Remazeilles, Renzi, Rocha, Rosset, Roudier, Rubi{\~n}o-Mart{\'\i}n, Ruiz-Granados, Salvati, Sandri, Savelainen, Scott, Shellard, Sirignano, Sirri, Spencer, Sunyaev, Suur-Uski, Tauber, Tavagnacco, Tenti, Toffolatti, Tomasi, Trombetti, Valenziano, Valiviita, Van~Tent, Vibert, Vielva, Villa, Vittorio, Wandelt, Wehus, White, White, Zacchei, \& Zonca}]{PlanckCollaboration2020}
{Planck Collaboration}, Aghanim, N., Akrami, Y., {et~al.} 2020, \aap, 641, A6

\bibitem[{Pritzl {et~al.}(2005)Pritzl, Venn, \& Irwin}]{Pritzl2005}
Pritzl, B.~J., Venn, K.~A., \& Irwin, M. 2005, \aj, 130, 2140

\bibitem[{Proctor {et~al.}(2004)Proctor, Forbes, \& Beasley}]{Proctor2004}
Proctor, R.~N., Forbes, D.~A., \& Beasley, M.~A. 2004, \mnras, 355, 1327

\bibitem[{Prugniel \& Soubiran(2004)}]{Prugniel2004}
Prugniel, P. \& Soubiran, C. 2004, arXiv e-prints, astro

\bibitem[{Riess {et~al.}(1998)Riess, Filippenko, Challis, Clocchiatti, Diercks, Garnavich, Gilliland, Hogan, Jha, Kirshner, Leibundgut, Phillips, Reiss, Schmidt, Schommer, Smith, Spyromilio, Stubbs, Suntzeff, \& Tonry}]{Riess1998}
Riess, A.~G., Filippenko, A.~V., Challis, P., {et~al.} 1998, \aj, 116, 1009

\bibitem[{Riess {et~al.}(2022)Riess, Yuan, Macri, Scolnic, Brout, Casertano, Jones, Murakami, Anand, Breuval, Brink, Filippenko, Hoffmann, Jha, D'arcy~Kenworthy, Mackenty, Stahl, \& Zheng}]{Riess2022}
Riess, A.~G., Yuan, W., Macri, L.~M., {et~al.} 2022, \apjl, 934, L7

\bibitem[{Roediger {et~al.}(2014)Roediger, Courteau, Graves, \& Schiavon}]{Roediger2014}
Roediger, J.~C., Courteau, S., Graves, G., \& Schiavon, R.~P. 2014, \apjs, 210, 10

\bibitem[{Salaris \& Cassisi(2005)}]{Salaris2005}
Salaris, M. \& Cassisi, S. 2005, Evolution of Stars and Stellar Populations

\bibitem[{Salim {et~al.}(2018)Salim, Boquien, \& Lee}]{Salim2018}
Salim, S., Boquien, M., \& Lee, J.~C. 2018, \apj, 859, 11

\bibitem[{Schiavon {et~al.}(2004)Schiavon, Rose, Courteau, \& MacArthur}]{Schiavon2004}
Schiavon, R.~P., Rose, J.~A., Courteau, S., \& MacArthur, L.~A. 2004, \apjl, 608, L33

\bibitem[{Schiavon {et~al.}(2005)Schiavon, Rose, Courteau, \& MacArthur}]{Schiavon2005}
Schiavon, R.~P., Rose, J.~A., Courteau, S., \& MacArthur, L.~A. 2005, \apjs, 160, 163

\bibitem[{Scott {et~al.}(2017)Scott, Brough, Croom, Davies, van~de Sande, Allen, Bland-Hawthorn, Bryant, Cortese, D'Eugenio, Federrath, Ferreras, Goodwin, Groves, Konstantopoulos, Lawrence, Medling, Moffett, Owers, Richards, Robotham, Tonini, \& Yi}]{Scott2017}
Scott, N., Brough, S., Croom, S.~M., {et~al.} 2017, \mnras, 472, 2833

\bibitem[{Soderblom(2010)}]{Soderblom2010}
Soderblom, D.~R. 2010, \araa, 48, 581

\bibitem[{Strader \& Brodie(2004)}]{Strader2004}
Strader, J. \& Brodie, J.~P. 2004, \aj, 128, 1671

\bibitem[{Thomas {et~al.}(2011)Thomas, Maraston, \& Johansson}]{Thomas2011}
Thomas, D., Maraston, C., \& Johansson, J. 2011, \mnras, 412, 2183

\bibitem[{Tomasetti {et~al.}(2023)Tomasetti, Moresco, Borghi, Jiao, Cimatti, Pozzetti, Carnall, McLure, \& Pentericci}]{Tomasetti2023}
Tomasetti, E., Moresco, M., Borghi, N., {et~al.} 2023, \aap, 679, A96

\bibitem[{Usher {et~al.}(2019{\natexlab{a}})Usher, Beckwith, Bellstedt, Alabi, Chevalier, Pastorello, Cerulo, Dalgleish, Fraser-McKelvie, Kamann, Penny, Foster, McDermid, Schiavon, \& Villaume}]{Usher2019a}
Usher, C., Beckwith, T., Bellstedt, S., {et~al.} 2019{\natexlab{a}}, \mnras, 482, 1275

\bibitem[{Usher {et~al.}(2019{\natexlab{b}})Usher, Brodie, Forbes, Romanowsky, Strader, Pfeffer, \& Bastian}]{Usher2019}
Usher, C., Brodie, J.~P., Forbes, D.~A., {et~al.} 2019{\natexlab{b}}, \mnras, 490, 491

\bibitem[{Usher {et~al.}(2017)Usher, Pastorello, Bellstedt, Alabi, Cerulo, Chevalier, Fraser-McKelvie, Penny, Foster, McDermid, Schiavon, \& Villaume}]{Usher2017}
Usher, C., Pastorello, N., Bellstedt, S., {et~al.} 2017, \mnras, 468, 3828

\bibitem[{Vagnozzi {et~al.}(2022)Vagnozzi, Pacucci, \& Loeb}]{Vagnozzi2022}
Vagnozzi, S., Pacucci, F., \& Loeb, A. 2022, Journal of High Energy Astrophysics, 36, 27

\bibitem[{Valcin {et~al.}(2020)Valcin, Bernal, Jimenez, Verde, \& Wandelt}]{Valcin2020}
Valcin, D., Bernal, J.~L., Jimenez, R., Verde, L., \& Wandelt, B.~D. 2020, \jcap, 2020, 002

\bibitem[{Valcin {et~al.}(2021)Valcin, Jimenez, Verde, Bernal, \& Wandelt}]{Valcin2021}
Valcin, D., Jimenez, R., Verde, L., Bernal, J.~L., \& Wandelt, B.~D. 2021, \jcap, 2021, 017

\bibitem[{van~den Bergh(1969)}]{Bergh1969}
van~den Bergh, S. 1969, \apjs, 19, 145

\bibitem[{van~den Bergh(1981)}]{Bergh1981}
van~den Bergh, S. 1981, \aaps, 46, 79

\bibitem[{VandenBerg {et~al.}(1996)VandenBerg, Bolte, \& Stetson}]{VandenBerg1996}
VandenBerg, D.~A., Bolte, M., \& Stetson, P.~B. 1996, Annual Review of Astronomy and Astrophysics, 34, 461

\bibitem[{Vazdekis {et~al.}(2015)Vazdekis, Coelho, Cassisi, Ricciardelli, Falc{\'o}n-Barroso, S{\'a}nchez-Bl{\'a}zquez, La~Barbera, Beasley, \& Pietrinferni}]{Vazdekis2015}
Vazdekis, A., Coelho, P., Cassisi, S., {et~al.} 2015, \mnras, 449, 1177

\bibitem[{Vazdekis {et~al.}(2010)Vazdekis, S{\'a}nchez-Bl{\'a}zquez, Falc{\'o}n-Barroso, Cenarro, Beasley, Cardiel, Gorgas, \& Peletier}]{Vazdekis2010}
Vazdekis, A., S{\'a}nchez-Bl{\'a}zquez, P., Falc{\'o}n-Barroso, J., {et~al.} 2010, \mnras, 404, 1639

\bibitem[{Walcher {et~al.}(2009)Walcher, Coelho, Gallazzi, \& Charlot}]{Walcher2009}
Walcher, C.~J., Coelho, P., Gallazzi, A., \& Charlot, S. 2009, \mnras, 398, L44

\end{thebibliography}

\end{document}